\documentclass[fleqn,usenatbib]{mnras}

\usepackage{newtxtext,newtxmath}
\usepackage{longtable}

\usepackage[T1]{fontenc}

\DeclareRobustCommand{\VAN}[3]{#2}
\let\VANthebibliography\thebibliography
\def\thebibliography{\DeclareRobustCommand{\VAN}[3]{##3}\VANthebibliography}

\usepackage{graphicx}	
\usepackage{amsmath}	
\usepackage{verbatim}

\newcommand{\chone}{$[3.6\mu{\rm m}]$}
\newcommand{\chtwo}{$[4.5\mu{\rm m}]$}

\newcommand{\cii}{{\sc [CII]~}}
\newcommand{\oiii}{{\sc [OIII]~}}
\newcommand{\oiiio}{{\sc [OIII]} $\lambda \lambda 4959,5007$~}
\newcommand{\as}{$\,{\rm arcsec}$}
\newcommand{\irxb}{${\rm IRX}$--$\beta$}

\newcommand{\muv}{${M_{\rm UV}}$}

\newcommand{\Lir}{$L_{\rm IR}$}
\newcommand{\Luv}{$L_{\rm UV}$}
\newcommand{\sfrunit}{$/{\rm M}_{\odot}/{\rm yr}$}
\newcommand{\mstar}{$M_{\star}$}
\newcommand{\lmstar}{${\rm log}_{10}(M_{\star}/{\rm M}_{\odot})$}
\newcommand{\irxm}{${\rm IRX}$--\mstar}

\def\be{\begin{equation}} 
\def\ee{\end{equation}}

\def\gsim{\lower.5ex\hbox{\gtsima}} 
\def\lsim{\lower.5ex\hbox{\ltsima}} \def\gtsima{$\; \buildrel > \over 
\sim \;$} \def\ltsima{$\; \buildrel < \over \sim \;$} \def\prosima{$\; 
\buildrel \propto \over \sim \;$} \def\gsim{\lower.5ex\hbox{\gtsima}} 
\def\lsim{\lower.5ex\hbox{\ltsima}} 
\def\simgt{\lower.5ex\hbox{\gtsima}} 
\def\simlt{\lower.5ex\hbox{\ltsima}} 
\def\simpr{\lower.5ex\hbox{\prosima}}   
  
 \def\gtsima{$\; \buildrel > \over \sim \;$} 
\def\ltsima{$\; \buildrel < \over \sim \;$} 
\def\gsim{\lower.5ex\hbox{\gtsima}} 
\def\lsim{\lower.5ex\hbox{\ltsima}} 
\def\simgt{\lower.5ex\hbox{\gtsima}} 
\def\simlt{\lower.5ex\hbox{\ltsima}} 
\def\simpr{\lower.5ex\hbox{\prosima}}

\def\E3{{\cal E}_{\rm g}^{III}}

\def\M*{M_*}

\def\Z*{Z_*}
\def\L*{L_*}

\title[REBELS: IRX-$\beta$ and IRX-{$M_{\star}$} relations]{The ALMA REBELS survey: obscured star formation in massive Lyman-break galaxies at $\mathbf{z = 4}$--$\mathbf{8}$ revealed by the IRX--{$\mathbf{\beta}$} and~{$\mathbf{M_{\star}}$}~relations}

\author[R. A. A. Bowler et al.]{R. A. A. Bowler,$^{1}$\thanks{rebecca.bowler@manchester.ac.uk} 
H. Inami,$^{2}$
L. Sommovigo,$^{3}$
R. Smit,$^{4}$
H. S. B. Algera,$^{2,5}$
M. Aravena,$^{6}$
L. Barrufet,$^{7}$\newauthor
R. Bouwens,$^{8}$
E. da Cunha,$^{9,10}$
F. Cullen,$^{11}$
P. Dayal,$^{12}$
I. De Looze,$^{13, 14}$
J. S. Dunlop,$^{11}$
Y. Fudamoto,$^{5, 15}$\newauthor
V. Mauerhofer,$^{12}$
R. J. McLure,$^{11}$
M. Stefanon,$^{16}$
R. Schneider,$^{17,18,19,20}$
A. Ferrara,$^{3}$
L. Graziani,$^{17,18}$\newauthor
J. A. Hodge,$^{8}$
T. Nanayakkara,$^{21}$
M. Palla,$^{13}$
S. Schouws,$^{8}$
D. P. Stark,$^{22}$
P. P. van der Werf$\,^{8}$\\
\vspace{1cm}
{\it \normalsize Affiliations are listed at the end of the paper}
}

\date{Accepted XXX. Received YYY; in original form ZZZ}

\pubyear{2023}

\begin{document}
\label{firstpage}
\pagerange{\pageref{firstpage}--\pageref{lastpage}}
\maketitle

\begin{abstract}
We investigate the degree of dust obscured star formation in 49 massive (\lmstar$>9$) Lyman-break galaxies (LBGs) at $z = 6.5$--$8$ observed as part of the ALMA Reionization Era Bright Emission Line Survey (REBELS) large program.
By creating deep stacks of the photometric data and the REBELS ALMA measurements we determine the average rest-frame UV, optical and far-infrared (FIR) properties which reveal a significant fraction ($f_{\rm obs} = 0.4$--$0.7$) of obscured star formation, consistent with previous studies.
From measurements of the rest-frame UV slope, we find that the brightest LBGs at these redshifts show bluer ($\beta \simeq -2.2$) colours than expected from an extrapolation of the colour-magnitude relation found at fainter magnitudes.
Assuming a modified blackbody spectral-energy distribution (SED) in the FIR (with dust temperature of $T_{\rm d} = 46\,{\rm K}$ and $\beta_{\rm d} = 2.0$), we find that the REBELS sources are in agreement with the local ``Calzetti-like'' starburst Infrared-excess (IRX)-$\beta$~relation.
By reanalysing the data available for 108 galaxies at $z \simeq 4$--$6$ from the ALPINE ALMA large program using a consistent methodology and assumed FIR SED, we show that from $z \simeq 4$--$8$, massive galaxies selected in the rest-frame UV have no appreciable evolution in their derived~\irxb~relation.
When comparing the~\irxm~relation derived from the combined ALPINE and REBELS sample to relations established at $z < 4$, we find a deficit in the IRX, indicating that at $z > 4$ the proportion of obscured star formation is lower by a factor of $\gtrsim 3$ at a given a \mstar.
Our \irxb~results are in good agreement with the high-redshift predictions of simulations and semi-analytic models for $z \simeq 7$ galaxies with similar stellar masses and SFRs.
\end{abstract}




\section{Introduction}

The onset of dust creation represents a milestone in the history of the Universe, as it relies on the adequate enrichment of the galaxy with metals, formation of sufficient dust particles in high-redshift supernova and inter-stellar medium (ISM) properties conducive to the survival (and growth) of dust (e.g.~\citealp{Draine03, Mancini16, Gall18, Lesniewska19, Graziani20, Dayal22, Cesare23}).
The presence of dust within galaxies can be detected through the reddening of the rest-frame UV and optical light in addition to emission in the mid and far-infrared (FIR).
The measurement of the rest-frame FIR modified blackbody emission provides a direct signal of the presence of dust, whereas changes in the rest-frame UV colours can be attributed to other properties of the galaxy such as older ages and an increased metallicity.
The majority of the highest-redshift galaxies found within deep optical to near-infrared (NIR) surveys have been shown to have blue rest-frame UV slopes (parameterised as $f_{\lambda} \propto \lambda^{\beta}$, $\beta \simeq -2$), leading to the inference of young ages and low dust content~\citep{Dunlop12, Bouwens14}.
In the past decade however, the direct detection of dust continuum emission in individual or small samples of $z \gtrsim 7$ galaxies (e.g.~\citealp{Tamura19, Wong22, Hygate23, Hashimoto23}) has revealed the presence of dust within galaxies less than 800 Myr after the Big Bang.

Observations of star-forming galaxies in the rest-frame FIR have demonstrated the key importance of considering dust obscured star-formation in galaxy evolution, with more than half of ongoing star formation being obscured at cosmic noon ($z \simeq 3$; see review by~\citealp{Madau14}).
There is evidence that obscured star formation continues to be important, and potentially dominates the total cosmic SFR density (CSFRD) in the range $3 < z < 6$, from measurements based on rest-frame UV selected samples (e.g.~\citealt{Novak17, Khusanova21}) and highly dust-obscured galaxies including serendipitous objects (e.g.~\citealp{Gruppioni20, Talia21, Loiacono21}), as well as deep ALMA and radio surveys (e.g.~\citealp{Zavala21, Vlugt22}).
Recent results extending these measurements to $z \simeq 7$ from~\citet{Barrufet23} have shown that dust obscured star-formation contributes at least 10 percent of the cosmic star-formation rate density, showing that it remains significant even into the Epoch of Reionization.
These results have revealed a strong stellar mass dependence of the obscuration (e.g.~\citealp{Pannella09, Pannella15, Bouwens16, Whitaker17}), with~\citet{Dunlop17} demonstrating that at $z \simeq 2$ the fraction of obscured SFR rises from $\lesssim 0.5$ at \lmstar $ < 9$ up to 0.99 at \lmstar $ > 10$, an effect which appears to extend to $z \simeq 7$ (although with a lower normalisation;~\citealp{Algera23}).
Direct detections of the dust continuum emission from galaxies at $z > 6.5$ have been made in galaxies selected from galaxies representing a wide range of intrinsic rest-frame UV luminosities for example fainter sources from lensing fields (e.g.~\citealp{Watson15, Laporte17, Tamura19, Bakx21, Hashimoto23}) and brighter galaxies from wide-area ground-based follow-up (e.g.~\citealt{Bowler18, Bowler22, Schouws22, Inami22, Witstok22}).
The obscured fraction derived from these works depends on the assumed FIR spectral-energy distribution (e.g. typically the dust temperature; $T_{\rm d}$ and the emissivity index; $\beta_{\rm d}$), however in general these detections reveal a significant fraction $\simeq 0.2$--$0.8$ of obscured star-formation at $z \simeq 7$ for galaxies of \lmstar$ \simeq 9.5$~\citep{Dayal22, Algera23}.

The attenuation curve, which dictates how an intrinsic spectrum is reduced in the rest-frame UV and optical as a function of wavelength for a given optical depth, depends on the detailed properties of the dust grains and their geometric distribution (see~\citealp{Salim20} for a review).
To directly measure the attenuation curve requires a handle on the intrinsic stellar spectra before the effect of dust, a technique that has been successfully employed at $z =2$--$5$ (e.g.~\citealp{Cullen18,Shivaei20}).
An alternative method is to compare the rest-frame UV slope, $\beta$, to the ratio of the FIR to UV luminosity (infrared-excess $= IRX = {\rm log}_{\rm 10}(L_{\rm FIR}/L_{\rm UV})$) as the steepness of the attenuation (or extinction) curve changes the relation between IRX and $\beta$ to maintain energy balance.
The so called ``\irxb'' diagram for local starburst galaxies shows a strong correlation presented originally in~\citet{Meurer99} and then further refined in~\citet{Calzetti00}.
Whether this canonical Calzetti-relation holds at higher redshifts ($z \gtrsim 2$) has been the topic of debate over the past decade.
An alternative to the Calzetti-like attenuation curve is the steeper relation that has been found for the Small Magellanic Cloud (SMC).
Note that the SMC relation is an~\emph{extinction} as opposed to an~\emph{attenuation} curve, and there is an ongoing discussion on whether it is expected that observations of galaxies will be consistent with an SMC extinction curve when the likely complex geometry of dust is taken into account (e.g. see discussion in~\citet{Cullen18}.
In this case, the same column density of dust can provide an increased reddening effect in the rest-frame UV and hence a deficit from the Calzetti-like relation.
Initial observations of $z = 5$--$6$ galaxies with ALMA suggested such a deficit was found (e.g.~\citealp{Capak15, Barisic17}), however other studies~\citep{Bowler18, Bowler22, Schouws22} found results consistent with the Calzetti-like \irxb.
Note that the discrepancy between these studies is reduced if we consider that the first works typically assumed dust temperatures of $T_{\rm d} = 25$--$45\,{\rm K}$, while later works tended to use higher temperatures ($T_{\rm d} \simeq 50\,{\rm K}$).
However the observation of several galaxies at $z \simeq 5$ that lie below the SMC prediction is still present with higher temperatures as shown by~\citealp{Faisst17}.
Furthermore, individual sources at $z \simeq 7$ have been found to show significant scatter both above and below a Calzetti-like relation~\citep{Smit18, Hashimoto19, Bakx20}, while fainter (and likely lower mass) sources appear to show an upper limit that is even below the prediction of an SMC-like extinction curve~\citep{Fujimoto16, Bouwens16}.

One key uncertainty in the measurement of the \irxb~and~\irxm~relations at increasingly high redshifts is that even at $z \gtrsim 2$ there are very few individual detections of dust continuum emission from galaxies at \lmstar$< 10$ (e.g.~\citealp{Dunlop17, Bouwens20}).
Because of this many studies at $z \gtrsim 2$ rely on stacking analyses of large numbers of individually undetected galaxies within rest-frame FIR survey data (e.g. from SCUBA-2;~\citealp{Koprowski18}) or alternatively small samples of often inhomogeneously detected samples from multiple follow-up programs with ALMA.
In addition to the reliance on stacking or small samples, there are several systematic uncertainties that have precluded a deeper understanding of the~\irxb~relation at high redshift.
The first is the uncertain FIR Spectral-energy distribution (SED) in the galaxies of interest, as the majority of early studies rely on measurements in the rest-frame FIR at typically one frequency.
The derived FIR luminosity is strongly dependent on the assumed SED ($L_{\rm IR} \propto T_{\rm d}^{4+\beta_{\rm d}}$; see discussion in e.g.~\citealp{Behrens18, Liang19, Sommovigo20}) and where individual dust temperature measurements have been made a wide variation has been found in the derived $T_{\rm d} = 20$--$70\,{\rm K}$~\citep{Witstok23}.
Second, the position of a galaxy with respect to the~\irxb~relation depends sensitively on the geometry of the dust and stars, as has been shown in theoretical works (e.g.~\citealp{Popping17, Narayanan18, Ferrara22, Pallottini22, Vijayan23}).
Early resolved observations have shown that there appears to be an anti-correlation between the position of the rest-frame UV and FIR emission, suggesting a complex geometry that could impact the observed~\irxb~\citep{Faisst17, Carniani17, Bowler18, Inami22, Hashimoto23, Tamura23}.

The result of these studies is an uncertain picture of how the commonly observed rest-frame UV emission in LBGs is connected to any obscured star formation at $z \gtrsim 7$ (see~\citealp{Hodge20} for a recent summary).
To make progress in understanding obscured star formation in LBGs what is required is a statistical survey of homogeneously selected galaxies with deep observations probing the dust continuum.
In this work we utilise a comprehensive survey of 49 rare, massive (\lmstar $\gtrsim 9$) galaxies at $z = 6.5$--$8.5$ observed as part of the ALMA REBELS large program~\citep{Bouwens22}.
The majority of these sources were selected from wide-area, ground-based data over $7\,{\rm deg}^2$ and they probe bright rest-frame UV magnitudes $M_{\rm UV} < -21$ and hence the bright-end of the rest-frame UV luminosity function at this epoch (e.g.~\citealt{Bowler17, Harikane22, Varadaraj23}).
We also perform a consistent analysis of the ALMA ALPINE large program~\citep{LeFevre20, Bethermin20, Faisst20} to provide a measurement of the evolving~\irxb~and~\irxm~relations from $z = 4$ to $ z = 8$ using the most comprehensive homogeneous samples of $z > 4$ galaxies observed with ALMA.
REBELS provides a unique sample of galaxies with direct dust detections (or strong upper limits) to constrain the \irxb~and \irxm~relation within the Epoch of Reionization (EoR).
This work builds upon the previous observational REBELS papers from~\citet{Inami22}, \citet{Algera23}, and~\citet{Barrufet23} and theoretical analyses tailored specifically to describe the REBELS observations from~\citet{Dayal22},~\citet{Sommovigo22a} and~\citet{Ferrara22}.

The structure of this paper is as follows.
In Section~\ref{sect:data} we describe our sample from REBELS and ALPINE, presenting the ALMA observations in addition to the archival optical and NIR data that we utilize.
We present the methods and results in Section~\ref{sect:methods}, in particular the stacking analysis and SED fitting.
In Section~\ref{sect:results} we present the resulting colour-magnitude relation, physical properties from SED fitting and the~\irxb~relation.
In Section~\ref{sect:disc} we discuss our results and present a new derivation of the~\irxm~relation from $z = 4$--$8$, and we compare our ALPINE + REBELS results to the predictions from simulations in Section~\ref{sect:sim}.
We end with our conclusions in Section~\ref{sect:conc}.
Throughout this work we present magnitudes in the AB system~\citep{Oke83}.
The standard concordance cosmology~\citep{Aghanim20} is assumed, with $H_{0} = 70 \, {\rm km}\,{\rm s}^{-1}\,{\rm Mpc}^{-1}$, $\Omega_{\rm m} = 0.3$ and $\Omega_{\Lambda} = 0.7$.

\section{Data}\label{sect:data}
In this work we combine rest-frame UV, optical and FIR measurements to understand the dust properties of $z =4$--$8$ galaxies.
The rest-frame UV and optical information is provided by deep degree-scale extragalactic survey observations that have the required wavelength coverage from multiple photometric bands to select these galaxies via the redshifted Lyman break.
The rest-frame FIR measurements come from targeted ALMA programs to follow-up these bright sources and provide a direct detection or upper limit on the dust-continuum emission.

\subsection{REBELS}
The REBELS survey is a Cycle 7 ALMA large program that observed 40 LBGs with the primary goal to measure the \cii $158\,\mu{\rm m}$ or \oiii $88\,\mu{\rm m}$ emission line.
The sources were found within the ground-based Cosmological Evolution Survey (COSMOS;~\citealp{Scoville07}) and the~\emph{XMM-Newton} Large Scale Structure (XMM-LSS;~\citealp{Pierre04}) surveys, with the addition of two sources from~\emph{HST} surveys (REBELS-16 and REBELS-40).
We also include 9 sources that were observed as part of the REBELS pilot programs as presented in~\citet{Smit18} and~\citet{Schouws22}, resulting in a final sample containing 49 galaxies.
The primary selection criterion was that the source redshift lay securely at $z > 6.5$ as determined by three independent SED fitting codes.
The galaxies are bright in the rest-frame UV, with absolute magnitudes (measured at $1500$\AA~in the rest-frame) in the range $-23.0 < M_{\rm UV} < -21.3$.
REBELS observed each source using between two and six spectral tunings to cover the frequency range of likely FIR line emission given the photometric redshift probability distribution.
As presented in~\citet{Bouwens22}, \citet{Inami22}, Schouws et al. (2023, in prep), and van Leeuwen et al. (2023, in prep), 25 galaxies have been spectroscopically confirmed via the \cii line (with no \oiii confirmations to-date).
In addition, these observations simultaneously allowed a measurement of the dust-continuum emission, with 18 sources detected in the continuum at $> 3.3\sigma$ by~\citet{Inami22}.
The \cii line (if detected) was masked in the continuum images.
The typical continuum depth of the Band 6 or 7 data (approximately $240\,{\rm GHz}$ and $350\,{\rm GHz}$ depending on exact line scan frequencies) was $\sigma_{\rm rms} = 10$--$20\,\mu {\rm Jy}$ with a beam of $1.2$--$1.6''$ full width at half maximum (FWHM).

\subsection{ALPINE}
The ALPINE survey is an ALMA large program awarded in Cycle 5 that aimed at detecting the \cii~line and dust continuum emission in 118 galaxies at $z = 4$--$6$~\citep{LeFevre20, Bethermin20}.
The sources were selected from a large red-optical spectroscopic survey of ``normal'' star-forming galaxies within the COSMOS and Extended-\emph{Chandra} Deep Field South (ECDFS) fields.
The resulting sample consisted of 67 sources with spectroscopic redshifts from Lyman$-\alpha$ emission or rest-frame UV absorption features between $z = 4.4$--$4.6$ and 51 objects at $z = 5.1$--$5.9$~\citep{Faisst20a}.
Both Lyman-break selection and narrow-band selections were utilized (with more narrow-band sources in the $z > 5$ sub-sample), leading to a relatively high average rest-frame EW of Ly$\alpha$ of $\simeq 5$--$100$\AA~\citep{Cassata20}.
In comparison to REBELS, where none of the sources had spectroscopic redshifts prior to the ALMA program, this leads to a different selection function for galaxies, which we discuss further in Section~\ref{sect:disc}.
As presented in~\citet{Fudamoto20}, 23 galaxies were detected at $> 3.5\sigma$ significance in the dust continuum in the original ALPINE survey (8 sources at $z > 5$).
Several sources have been further followed-up in multiple bands (e.g. HZ4 and HZ6; see~\citealp{Faisst20}).
The typical depth of the ALPINE Band 7 data ($275$--$373\,{\rm GHz}$) was $\sigma_{\rm rms} = 30 (50)\,\mu {\rm Jy}$ for the $z = 5.5$ ($4.5$) samples, with an average beam of $1.1''$ FWHM~\citep{Bethermin20}.

\subsection{Optical and near-infrared imaging}\label{sect:cats}
To measure the rest-frame UV slopes of the REBELS and ALPINE galaxies, in addition to physical properties such as stellar mass, we exploited the wealth of available optical and NIR imaging data.
The details of the photometry for the REBELS sample is presented in~\citet{Bouwens22} and Stefanon et al. (in prep), however we briefly describe the relevant data here.
In the COSMOS (XMM-LSS) field we used the UltraVISTA (VIDEO) survey from VISTA which provided imaging in the NIR $YJHK_{s}$-bands~\citep{McCracken12, Jarvis13}.
A subset of (fainter) galaxies were additionally located within a $1\,{\rm deg}^2$ sub-region of the XMM-LSS field that has deeper observations in the $JHK$ from the UK Infrared Deep Sky Survey~\citep{Lawrence07} Ultra-Deep Survey (UDS).
\emph{Spitzer}/Infrared Array Camera (IRAC) photometry was extracted from the deep mosaics presented in Stefanon et al. (in prep.), in particular from the~\emph{Spitzer} Extended Deep Survey (SEDS;~\citealp{Ashby13}) and the~\emph{Spitzer} Matching Survey of the UltraVISTA Ultra-deep Stripes (SMUVS;~\citealp{Ashby18}).
Photometry was extracted using $0.6''$ diameter apertures ($0.9''$ for IRAC) on images where the neighbouring sources had been subtracted using {\sc MOPHONGO}~\citep{Labbe15}.
The aperture flux was corrected to total according to the {\sc MOPHONGO} model for the galaxy.
As several of the very bright ($M_{\rm UV} <-22.5$) REBELS sources were resolved in the ground-based data~\citep{Bowler17}, this step was important in deriving accurate absolute magnitudes and physical properties.
Errors on the photometry were derived from empty aperture measurements on the data.

\begin{figure}
    \centering
    \includegraphics[width = 0.48\textwidth]{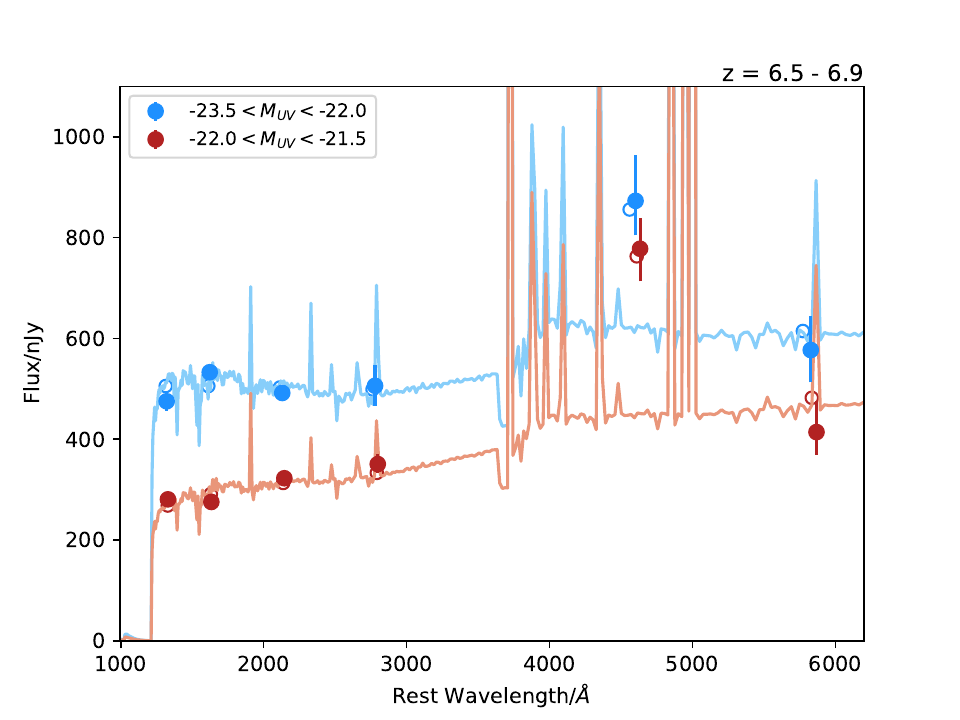}\\
    \includegraphics[width = 0.48\textwidth]{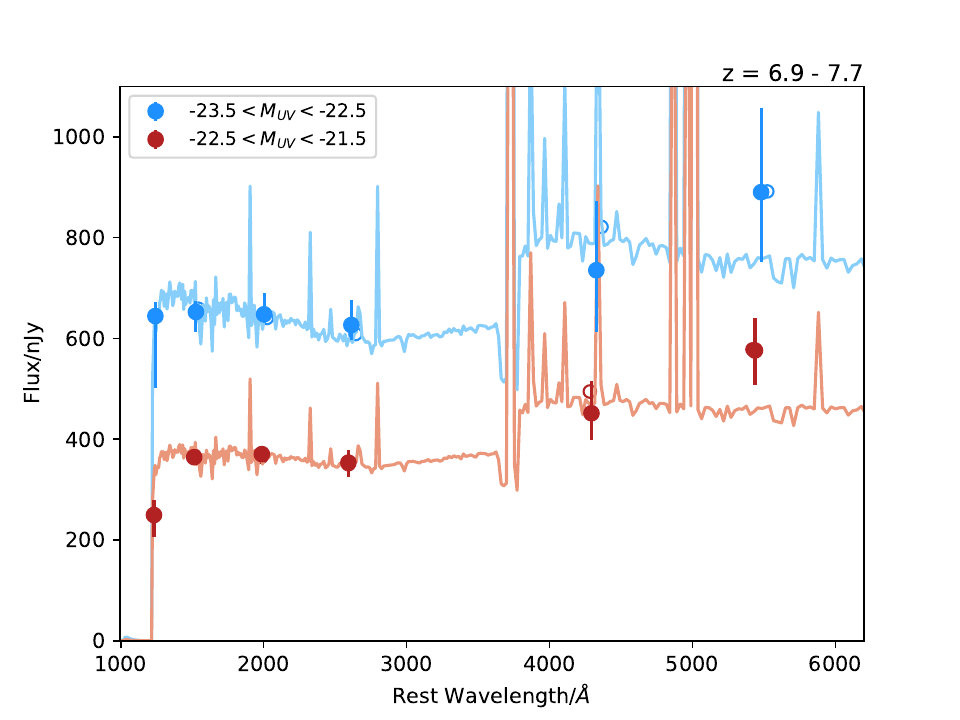}
    \caption{The average optical and NIR photometry and corresponding best-fitting SED model for the REBELS samples at $z < 6.9$ (top) and $ z > 6.9$ (bottom).
    In each plot we show the observed fluxes and errors from the bright and faint stacks as the blue and red filled circles respectively.
    The six points correspond to the VISTA $YJHK_{s}$ and~\emph{Spitzer}/IRAC \chone~and \chtwo~bands.
    The open circles correspond to the synthetic photometry as derived from the best-fitting SED model from {\sc BAGPIPES}, shown as the solid line.
    The observed change in the \chone~to \chtwo~colour with redshift is due to the transit of the strong H$\beta$+{\sc [OIII]} emission lines between these filters (e.g.~\citealp{Smit14}).}
    \label{fig:sed}
\end{figure}

For the ALPINE sample we used the photometry provided in the COSMOS2020 catalogue~\citep{Weaver22} which provided point-spread function (PSF) matched flux measurements for the full suite of optical and NIR filters available in the field.
In comparison to the COSMOS2015 catalogue that was utilised in the original ALPINE papers (as presented in~\citealt{Faisst20a}), the COSMOS2020 catalogue has deeper data in a range of filters.
Particularly for this work, the optical (from Subaru) and near-infrared (from UltraVISTA DR4) data are up to 1 magnitude deeper, providing significantly improved measurements of the rest-frame UV slope and derived stellar masses.
We used the `Classic' catalogue that provides photometry measured in $2''$ diameter apertures.
In the COSMOS2020 catalogue this aperture photometry is corrected to a total flux by a constant factor determined from the {\sc Source Extractor} MAG\_AUTO.
From the full ALPINE catalogue of 118 sources, ten sources lie within the ECDFS.
To provide a sample with uniform photometry from the COSMOS2020 catalogue, we excluded these ten sources from further analysis, leaving a final ALPINE sample of 108 galaxies.

\section{Methods}\label{sect:methods}
The primary goal of this work is to measure the rest-frame UV and FIR properties of the 49 bright $z = 6$--$8$ LBGs observed as part of the REBELS survey.
We also include for comparison a consistent analysis of the ALPINE sample in the COSMOS field, to provide a base-line reaching down to $z \simeq 4$.
We measure the properties of individual sources from both surveys, but also perform a stacking analysis to derive average properties within bins of \mstar~and \muv. 
Due to the relatively low fraction of sources that are directly detected in the dust continuum (0.43 for REBELS, 0.19 for ALPINE), stacking is a key tool to understand the dust continuum properties of galaxies within this sample.
Here we describe the key methods used in this work.
Where possible we used identical approaches for the REBELS and ALPINE analysis to provide a direct comparison between galaxies spanning $z \simeq 4$ to $z \simeq 8$.

\subsection{Stacking analysis}\label{sect:stacking}
The ALPINE and REBELS samples span a range of redshifts leading to different rest-frame features being observed in our available optical and NIR data.
We therefore separate our sample into two redshift bins in ALPINE and three within REBELS.
The two ALPINE bins correspond to $z = 4.0$--$4.5$ and $z = 4.8$--$5.4$, which are the two main redshift groupings within the sample.
All ALPINE sources were observed in the same ALMA band (Band 7).
In REBELS we split the sample into three bins, by increasing redshift as detailed below. 
We excluded the~\emph{HST} selected sources REBELS-16 and REBELS-40 from the stacks as they have different rest-frame optical and NIR filters available.
The first bin included galaxies at $z = 6.5$--$6.9$ (20 galaxies), with the second bin included galaxies in the range $z = 6.9$--$7.7$ (20 galaxies).
This separation at $z = 6.9$ was chosen as it is the point at which the Lyman break starts to move into the VISTA $Y$-band and is also when the H$\beta$+\oiiio rest-frame optical lines move from the \chone~to the \chtwo~band.
The 40 sources within these two bins had observations in ALMA Band 6.
The third and final REBELS bin included the seven galaxies that have $z > 7.7$.
Four of these sources have ALMA Band 7 observations that were designed to target the \oiii line.
These sources have photometric redshifts in the range $z = 7.7$--$8.6$, however none have been spectroscopically confirmed to-date.
Due to the wide range of photometric redshifts in this bin, and the differing ALMA measurement bands, we do not create a stack from this sub-sample.
However, we present their individual \irxb~properties for comparison with our stacks.
We further choose to split the sample using $M_{\rm UV}$ as this provided the greatest dynamic range in the derivation of the \irxb~and \irxm~relation, while also not suffering from biases (as the $\beta$ and \mstar~values have significant statistical errors, scatter between bins can lead to biases; see e.g.~\citealt{McLure18}).
The bins we used for REBELS are shown in Table~\ref{table:flux}.
We split the sub-samples by \muv$ \,= -22.0$ (\muv$\, = -22.5$) at $z \simeq 6.7$ ($z \simeq 7.2$) to provide roughly equal sources with the brighter/fainter absolute magnitude bins.
For ALPINE we split by \muv$ \,= -22.0$ for both redshift bins, and in addition we separate into two stellar mass bins to take into account the wider range of \mstar~values within the sample.
The ALPINE bins are detailed in Table~\ref{tab:alpine_flux}.
By restricting the \muv~and \mstar~ranges slightly, we result in a final ALPINE sample of 54 galaxies within the $z \simeq 4.5$ bin and 32 within the $z \simeq 5.5$ bin.

We performed weighted mean stacking of the ALMA continuum data in the specified bins in the image plane.
Our results are unchanged if we instead use a median stacking procedure.
Due to the majority of sources in the REBELS and ALPINE samples being undetected in the ALMA data we stack at the position of the observed rest-frame UV.
We note that this could cause an underestimate of the peak ALMA flux if there exists significant offsets between the rest-frame UV and FIR emission (e.g. as simulated in~\citealp{Bowler18}).
Given the large beam of the REBELS and ALPINE observations, and the relatively small absolute offsets found for these samples (e.g.~\citealp{LeFevre20, Inami22}, we expect the effect to be small.
Indeed, as described in Section~\ref{sect:fir}, only the most massive ALPINE stack shows evidence for extended emission, which we attribute partially to offsets between the rest-frame UV and FIR positions.
We take this into account in the flux measurement by using a Gaussian fit to the stack.

Foreground sources were masked based on a map of objects derived from {\sc PyBDSF}~\citep{Mohan15}, excluding pixels within a radius of $1.5$\arcsec~from the rest-frame UV centroid of the galaxy.
The majority of sources did not have any neighbours at the depth of the ALMA imaging. 
Errors on the stacked flux measurements were determined by remaking each stack using bootstrap resampling with replacement.
To determine the mean optical and NIR photometric measurements we averaged the fluxes from the REBELS and COSMOS2020 catalogues (Section~\ref{sect:cats}).
We also experimented with stacking the images themselves, however we concluded that it was not possible to improve on the flux stacking results, due to close neighbours contaminating the flux measurements.
This contamination was taken into account in our catalogue creation, with nearby sources being subtracted prior to aperture photometry (see Section~\ref{sect:cats}).
For ALPINE we use the COSMOS2020 `Classic' aperture photometry measurements where basic subtraction of neighbours is performed.
We visually checked the ALPINE sources in the COSMOS optical and NIR imaging, but found that they are all sufficiently isolated for the catalogue fluxes to be robust.

\subsection{SED fitting}\label{sect:bag}
We fit the photometric data for the individual REBELS and ALPINE sources (and the derived stacks) using {\sc BAGPIPES}~\citep{Carnall18} to provide a best-fitting model with which to measure the rest-frame UV slope.
The fitting also provides physical properties for the stacks, which we include in particular for measuring the~\irxm~relation.
We fix the redshift to the spectroscopic redshift when available (28 sources in REBELS, all of the sources in ALPINE), and for the stacked photometry we fix the redshift to the average redshift.
We found that using a luminosity-weighted redshift instead of an average had no effect on our results, as the difference was $\delta z \le 0.03$.
We include bands above the Lyman-break reaching to the \chtwo~filter, beyond which the resolution and depth decreases dramatically.
We also exclude bands that contain the Lyman-break in the fitting of the stacked photometry, as the small differences in break position within the band lead to tensions within the fitting.
Hence for REBELS, we fit to the $YJHK_{s}$\chone\chtwo~bands for the $z = 6.5$--$6.9$ sub-sample stack, and to the $JHK_{s}$\chone\chtwo~bands for the $z > 6.9$ stack.
The resulting photometry and best-fitting SED models for the REBELS stacks are shown in Fig.~\ref{fig:sed}.
For ALPINE, we fit to the $IzYJHK_{s}$\chone\chtwo bands for the $z = 4.5$ stack, and to the $zYJHK_{s}$\chone\chtwo bands for the $z = 5.5$ stack.
A delayed-$\tau$ model was assumed ($\Phi(t) \propto t e^{-(t/\tau)}$) in which the timescale of the decline was allowed to vary in the range $\tau = [0.3, 10.0]\,{\rm Gyr}$ and the age from $ 1\,{\rm Myr}$ up to the age of the Universe at that redshift.
The metallicity was fixed to $0.2\,{\rm Z}_{\odot}$, and the~\citet{Calzetti00} dust law was assumed with the attenuation in the $V$-band constrained to the range $A_{\rm V} = [0, 2]$.
We allowed the nebular ionization parameter to vary in the range ${\rm log}_{10}(U) = [-2, -4]$.
Uniform priors were assumed for all of the fitted parameters.
These parameters resulted in acceptable fits to the REBELS sources as seen in Fig.~\ref{fig:sed}, with no evidence for truncation of the resulting corner plots. 
Assuming a different SFH (e.g. constant or $\tau$) or metallicity only marginally affected the \mstar~by at most $0.1\,{\rm dex}$ and the derived $\beta$ values by $< 0.05$.
Note that assuming a non-parametric SFH for the REBEL sample as presented in~\citet{Topping22} can increase the derived stellar masses by on average $ \simeq 0.5\,{\rm dex}$ and in some cases $\gtrsim 1.5\,{\rm dex}$.
To provide a closer comparison to previous literature measurements of the~\irxm~we primarily consider the \mstar~values derived with standard parametric SFHs, however we note where relevant how our results would change with an assumed alternate SFH.
For the ALPINE sample,~\citet{Faisst20} found that the $\beta$-value derived depended on the assumed dust law in the fitting.
We also recover this trend in our sample, with the dervied $\beta$-slopes being redder by around $0.1$ when fitting with an SMC dust law in comparison to a Calzetti law.

\begin{figure}
    \centering
    \includegraphics[width=0.23\textwidth, trim = 0.2cm 0 0.2cm 0]{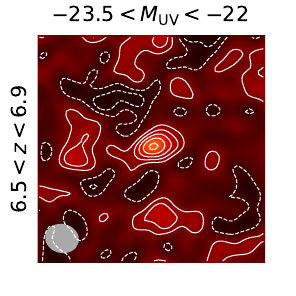}
     \includegraphics[width=0.23\textwidth, trim = 0.2cm 0 0.2cm 0]{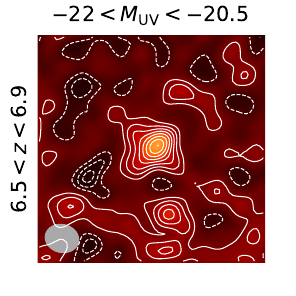}\\
    \includegraphics[width=0.23\textwidth, trim = 0.2cm 0 0.2cm 0]{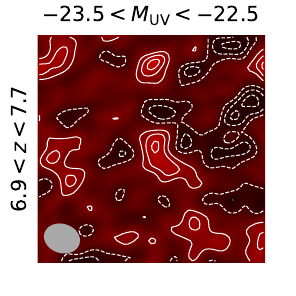}
    \includegraphics[width=0.23\textwidth, trim = 0.2cm 0 0.2cm 0]{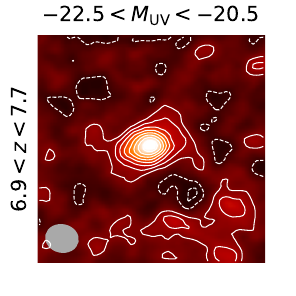}
    \caption{The ALMA Band 6 ($\lambda_{\rm rest} \simeq 150\,\mu {\rm m}$) stacks for REBELS.
    The stacks in the redshift range $6.5 < z < 6.9$ ($6.9 < z < 7.7$) are shown in the upper (lower) row.
    The left and right columns show the brighter and fainter $M_{\rm UV}$ stacks respectively, with the brighter stacks containing 7 (6) sources and the fainter stack containing 13 (14) sources at $z \simeq 6.7$ ($z \simeq 7.2$).
    The stamps are 10 \as~on a side, with N to the top and E to the left. 
    The colour scale is saturated beyond the range $[-2.0, 10.0]\sigma$ and contours are shown at $1\sigma$ intervals.
    The average beam for the data included in the stack is shown as the grey ellipse, with the position angle determined as the mode of the input image values. 
    The stacks are consistent with being unresolved at the resolution of the data ($1.2$--$1.6$ arcsec FWHM).
    }
    \label{fig:fullstack}
\end{figure}

\begin{table*}
\caption{The measured FIR fluxes and derived properties of the four REBELS stacks at $z = 6.5$--$7.7$.
The equivalent results for the ALPINE analysis is presented in Table~\ref{tab:alpine_flux}.
The top (bottom) two rows show the lower (higher) redshift stack, with the stacks ordered by $M_{\rm UV}$.
Columns 1 and 2 detail the redshift and number of sources included in each stack.
The average redshift and $M_{\rm UV}$ of each stack are shown in Columns 3 and 4.
Column 5 presents the measured peak ALMA flux, with the corresponding S/N shown in brackets.
The flux measured using a Gaussian fit is shown in Column 6.
The derived FIR luminosity (assuming $T_{\rm d} = 46\,{\rm K}$, $\beta_{\rm d} = 2.0$) and the resulting IRX value are shown in Columns 7 and 8.
The \Lir~was determined from the peak flux for the REBELS results.
Finally the rest-frame UV slope $\beta$ is presented in Column 9, as measured from the best-fitting SED model.
}
\begin{tabular}{ccccccccc}
\hline
$z$ bin & N & $z_{\rm mean}$ & $M_{\rm UV}$ & $F_{\rm Peak}$ & $ F_{\rm Gauss}$ & \Lir & IRX & $\beta_{\rm SED}$ \\
& & & $/{\rm mag}$ & $/\mu\,{\rm Jy}$ & $/\mu\,{\rm Jy}$  & $/10^{11}\,{\rm L}_{\odot}$ & \\
\hline
6.5 < z < 6.9 & 7 & 6.76 & $-22.34 \pm 0.14$ & $ 30.3 \pm 7.9 \, (5.6)$ & $ 31.1 \pm 9.9 $ & $ 1.4 \pm 0.4$ & $ -0.11_{-0.18}^{+0.16}$ & $ -2
.10_{-0.14}^{+0.14}$ \\[1ex]
6.5 < z < 6.9 & 13 & 6.67 & $-21.71 \pm 0.12$ & $ 44.5 \pm 10.4 \, (6.5)$ & $ 64.0 \pm 15.3 $ & $ 2.1 \pm 0.5$ & $ 0.31_{-0.16}^{+0.14}$ & $
-1.79_{-0.08}^{+0.08}$ \\
\hline
6.9 < z < 7.7 & 6 & 7.11 & $-22.75 \pm 0.18$ & $ < 36.0\,(2.0)$ & $--$ & $ < 1.9 $ & $ < -0.14 $ & $ -2.19_{-0.10}^{+0.10}$ \\[1ex]
6.9 < z < 7.7 & 14 & 7.26 & $-21.97 \pm 0.25$ & $ 36.3 \pm 9.6 \, (10.1)$ & $ 56.0 \pm 8.5 $ & $ 2.0 \pm 0.5$ & $ 0.18_{-0.23}^{+0.22}$ & $ -
2.10_{-0.07}^{+0.07}$ \\
\hline
\end{tabular}\label{table:flux}
\end{table*}

\subsection{Rest-frame UV luminosity and slope determination}
The monochromatic rest-frame UV luminosity was derived at $1500\,$\AA~using a top-hat filter of width $100\,$\AA~applied to the best-fitting SED model from {\sc BAGPIPES} for both the REBELS and ALPINE stacked photometry.
We note that the aperture photometry for both the REBELS catalogue and COSMOS2020 have been corrected to a total flux accounting for the full extent of the galaxy, and hence the derived $M_{\rm UV}$ can be considered a total absolute magnitude.
We found a systematic offset brightwards of $\Delta M_{\rm UV} = -0.1\,{\rm mag}$ between the ALPINE absolute magnitudes presented in~\citet{Faisst20} and those found in our analysis, with some sources having considerable offsets reaching $> 0.5{\rm mag}$.
Further inspection reveals this to be due to the improved photometry between the COSMOS2015 and COSMOS2020 catalogues.
The rest-frame UV slope is historically defined from a series of windows in the continuum from $\lambda_{\rm rest} = 1268$--$2580\,$\AA~\citep{Calzetti94}.
Different methods for measuring $\beta$ from the available photometric data in high-redshift galaxies have been extensively discussed, including the fitting of a power law or a power law with a Lyman-break to the photometry directly or to the fit SED model (e.g.~\citealp{Dunlop12, Rogers13}).
In this work we measure $\beta$ from the best-fitting SED model derived from {\sc BAGPIPES} (see Section~\ref{sect:bag}), excluding regions that are outside of the Calzetti windows, to avoid strong absorption or emission features.
Errors were derived using a bootstrap analysis, where we restacked the photometry and re-fit using {\sc BAGPIPES}.
Comparing the derived $\beta$ values to those presented in the original ALPINE analysis~\citep{Faisst20}, with find on average a very mild bias to bluer slopes by $0.05$ in our analysis, when comparing results obtained by fitting assuming the same dust attenuation law.
The scatter between individual objects can be large (up to $\delta \beta = 0.5$), but is within the errors of the derived $\beta$ values.

\subsection{Rest-frame FIR luminosity derivation}\label{sect:fir}
Using {\sc PyBDSF}~\citep{Mohan15} we measured both the peak flux and flux derived from a Gaussian fit for the individual sources and the stacked ALMA data.
We found that our stacked results were consistent with being unresolved for~\lmstar$< 10$ (i.e. the full REBELS sample and low-mass sub-sample of ALPINE).
We define an unresolved source if the measured major and minor axes from {\sc PyBDSF} are consistent with the beam size within the $1\sigma$ error (again derived within {\sc PyBDSF}).
Hence we used the peak flux measure in these cases.
For the ALPINE sample at \lmstar$> 10$ we instead used the Gaussian flux measurement, as the derived sizes from {\sc PyBDSF} were significantly resolved.
From the single data-point in the observed mm-regime from ALMA in Band 6 for ALPINE and Band 6 or 7 for REBELS we determined the total FIR luminosity by assuming a modified blackbody SED.
We corrected for the effect of the Cosmic Microwave Background following~\citet{daCunha13}, which results in an increase in the \Lir~by 10 percent for the REBELS sample.
While some sources within the two surveys have been observed in multiple ALMA bands~\citep{Algera23a}, we choose here to provide a uniform measure of \Lir~from the single main band that is available for the full REBELS + ALPINE samples.
For the analysis presented in this work we assumed a single fixed dust temperature of $T = 46\,{\rm K}$, with an opacity fixed to $\beta_{\rm d} = 2.0$ (consistent with that recently measured by~\citealp{Witstok23}).
This dust temperature was derived by the model of~\citet{Sommovigo22a} and was used by~\citet{Inami22}.
When showing the \irxb~relation we illustrate with an arrow the uncertainty introduced by this assumption.
We keep the dust temperature constant between the ALPINE and REBELS analysis, following the $T_{\rm d}$ analysis presented in~\citet{Sommovigo22a, Sommovigo22} who found that the dust temperatures between the samples were consistent at $46\,{\rm K}$ despite the different redshifts.
This finding was not expected given that other studies have found a redshift evolution in $T_{\rm d}$, however the exact form of the relation is still under debate especially at $z > 4$ (e.g.~\citealp{Sommovigo22a, Witstok23, Jones23}.
For example the trend found by~\citet{Schreiber18} up to $z = 4$ would predict a change of around $5\,{\rm K}$ between the redshifts of the ALPINE and REBELS results.
As we discuss further in Section~\ref{sect:irxalpine} even this small temperature difference can have an appreciable effect on the derived IRX and hence \irxb~and~\irxm~relations.
The dust temperature constraints that exist for the individual ALPINE~\citep{Faisst17} and REBELS~\citep{Witstok22, Algera23a} galaxies are consistent with our chosen $T_{\rm d}$ within the (substantial) errors, and show best-fit values from $20\,{\rm K}$ to $90\,{\rm K}$.
Due to the lack of $T_{\rm d}$ measurements for the vast majority of our sample, we are unable to account for this in our analysis and leave it to a future work.

\section{Results}\label{sect:results}
In Fig.~\ref{fig:sed} we present the stacked photometry and best-fitting SED model for the REBELS sample, split into the two main redshift (and further two~\muv~bins) as shown in Table~\ref{table:flux}.
The results of stacking the ALMA data in these bins are shown in Fig.~\ref{fig:fullstack}.
We find a significant ($7$--$10\sigma$) detection in the fainter $M_{\rm UV}$-bin for both the $z \simeq 6.7$ and the $z \simeq 7.2$ stacks.
In the brighter stacks we find marginal detections in the dust continuum, at $4\sigma$ and $2.5\sigma$ for the $z \simeq 6.7$ and the $z \simeq 7.2$ stack respectively.
The fluxes we derive are consistent with that found in the independent analysis of the REBELS sample by~\citet{Algera23}, who used a Monte Carlo stacking analysis to measure a correlation of \Lir~with stellar mass.
From these ALMA detections we then proceeded to compute the \Lir~and combine this with the rest-frame UV information ($L_{\rm UV}$, rest-frame UV slope) and the stellar mass as derived from {\sc BAGPIPES} as detailed below.

\subsection{Physical properties}\label{sect:sfr}

The splitting of the bright REBELS sample into two redshift bins separated at $z = 6.9$ allows us to provide high-S/N stacks of the rest-frame UV and optical emission where the strong H$\beta$ + \oiiio~lines sit within a single \chone~or \chtwo~band.
As shown in Fig.~\ref{fig:sed} we find that the REBELS sources are blue in the rest-frame UV, as probed by the $YJHK_s$ bands, with strong \chone--\chtwo~colours evident.
As we move to $z > 6.9$ we see a change in the IRAC colour indicative of H$\beta$+\oiiio moving into the \chtwo-band.
The derived SED fitting parameters from this photometry including stellar mass and age are presented in Table~\ref{table:sed}.
As expected from the individually results for REBELS presented in~\citet{Bouwens22}, the galaxies are massive, with $M_{\star} \gtrsim 10^{9}\,{\rm M}_{\odot}$, and moderate ages of the order of $40$--$130\,{\rm Myrs}$.
The derived dust attenuation is relatively low, as expected from the fact that the sample is rest-frame UV selected and shows blue $\beta$ slopes (see Section~\ref{sect:cm}).
Our most significantly FIR detected stack has the reddest $\beta = -1.8$ and strongest $A_{\rm V} = 0.65 \pm 0.05$.
We measure the unobscured SFR from the rest-frame UV emission using the luminosity at 1500\AA~and the conversion of~\citet{Madau14} for a constant SFR in the previous 100 Myr and a fixed metallicity of $Z = 0.1\,{\rm Z}_{\odot}$.
The SFR from the FIR was derived using the conversion based on the same assumptions on the SFH from~\citet{Madau14}, which provides an identical calibration to that used in the previous REBELS work by~\citealp{Algera23}.
Both calibrations were adjusted to a~\citet{Chabrier03} initial mass function (IMF).
The total SFR measured as the sum of these two components (${\rm SFR}_{\rm UV} + {\rm SFR}_{\rm IR}$) is in good agreement with that derived from the SED fitting.

\begin{table*}
\caption{The physical properties of the REBELS sources as derived from the stacked photometry.
We employ both SFR calibrations and SED fitting using {\sc BAGPIPES}, where the best-fitting models are shown in Fig.~\ref{fig:sed}.
Each row corresponds to a different stack in redshift and absolute UV magnitude as shown in Columns 1 and 2 respectively.
In Columns 3 and 4 we present the SFR derived from the rest-frame UV and FIR (see Section~\ref{sect:sfr}), with Column 5 showing the obscured SFR fraction derived from these quantities.
Columns 6 to 10 show the SFR, \mstar, age, $A_{\rm V}$ and ionization parameter as derived from {\sc BAGPIPES} assuming a SFH following a delayed $\tau$ model and a fixed metallicity of $Z = 0.2\,{\rm Z}_{\odot}$.
}
\begin{tabular}{cccccccccc}
\hline
$z$ bin & $M_{\rm UV}$ bin & ${\rm SFR}_{\rm UV}$ & ${\rm SFR}_{\rm IR}$ & $f_{\rm obs}$ & ${\rm SFR}_{\rm SED}$ & \lmstar & Age & ${\rm A}_{V}$ & ${\rm log}_{10}(
U)$ \\
& & \sfrunit & \sfrunit & & \sfrunit & & /Myr & /mag & \\
\hline
6.5 < z < 6.9 & $-23.5< M_{\rm UV} < -22.0$ & $ 23_{-3}^{+3}$ & $ 17_{-4}^{+4}$ & $ 0.42 \pm 0.11$ & $42_{-16}^{+12}$ & $9.6_{-0.2}^{+0.2}$ &
 $70_{-40}^{+60}$ & $0.36_{-0.09}^{+0.07}$ & $-3.1_{-0.4}^{+0.4}$ \\[1ex]
6.5 < z < 6.9 & $-22.0< M_{\rm UV} < -20.5$ & $ 13_{-2}^{+2}$ & $ 25_{-5}^{+5}$ & $ 0.66 \pm 0.16$ & $30_{-12}^{+17}$ & $9.4_{-0.2}^{+0.2}$ &
 $40_{-20}^{+30}$ & $0.65_{-0.05}^{+0.05}$ & $-2.7_{-0.3}^{+0.4}$ \\
 \hline
6.9 < z < 7.7 & $-23.5< M_{\rm UV} < -22.5$ & $ 34_{-5}^{+6}$ & $ < 24 $ & $ < 0.41 $ & $48_{-12}^{+15}$ & $9.8_{-0.3}^{+0.2}$ & $130_{-80}^{
+130}$ & $0.21_{-0.12}^{+0.13}$ & $-3.0_{-0.6}^{+0.7}$ \\[1ex]
6.9 < z < 7.7 & $-22.5< M_{\rm UV} < -20.5$ & $ 17_{-3}^{+4}$ & $ 24_{-6}^{+6}$ & $ 0.59 \pm 0.20$ & $35_{-10}^{+10}$ & $9.6_{-0.2}^{+0.2}$ &
 $110_{-70}^{+110}$ & $0.31_{-0.13}^{+0.10}$ & $-2.9_{-0.6}^{+0.6}$ \\
\hline
\end{tabular}\label{table:sed}
\end{table*}

\subsection{Colour-magnitude relation at $\mathbf{z\simeq7}$}\label{sect:cm}

\begin{figure}
    \centering
    \includegraphics[width = 0.48\textwidth]{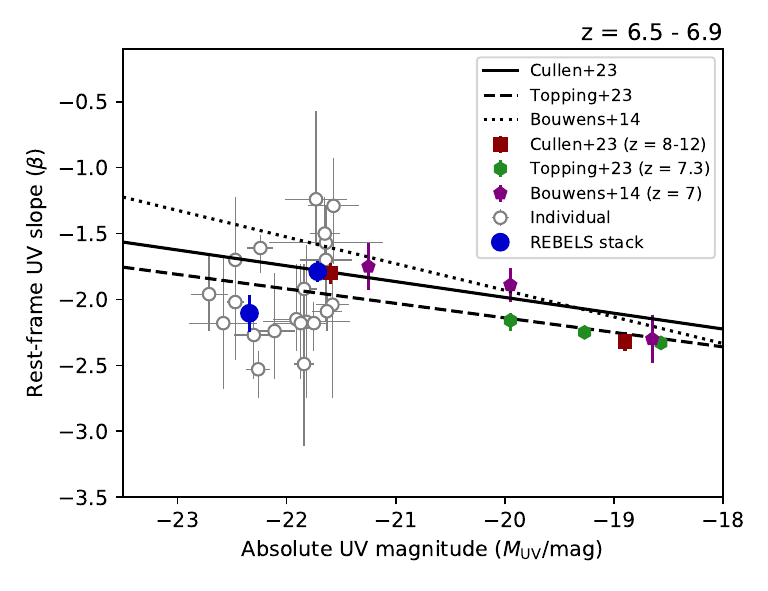}\\
    \includegraphics[width = 0.48\textwidth]{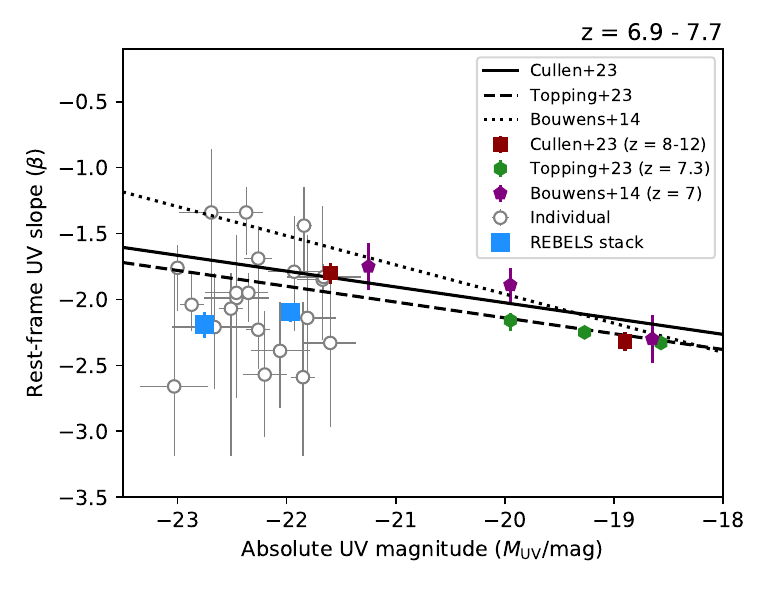}
    \caption{The REBELS galaxies at $z = 6.5$--$6.9$ (upper) and $6.9 < z < 7.7$ (lower) in comparison to the colour-magnitude relation found by previous studies.
    In each plot the individual galaxy measurements of the rest-frame UV slope and $M_{\rm UV}$ are shown as the grey open points, while the stacked results (from the fits shown in Fig.~\ref{fig:sed}) are shown as the blue filled points.
    The relationship derived from fainter studies are shown as the black solid, dotted and dashed lines from~\citet{Cullen23},~\citet{Topping23} and~\citet{Bouwens14} respectively.
    Slight differences between the relations in the upper and lower plot are due to the derived evolution of the relation from these studies.
    We also show the data points from the same three works (at a fixed rather than evolving redshift), note in particular that the redshift range of the data from the~\citet{Cullen23} study is at $z > 8$.
    }
    \label{fig:cm}
\end{figure}

In Fig.~\ref{fig:cm} we present the rest-frame UV slopes of the REBELS sources, and the stacks, in comparison to colour-magnitude relations from recent literature measurements.
The REBELS sample allows us to measure the colour-magnitude relation up to $M_{\rm UV} \simeq -23$, which is considerably brighter than the typical galaxies found and studied previously using~\emph{HST} or~\emph{JWST} data that are typically dominated by sources at $M_{\rm UV} \gtrsim -21$.
We find that the REBELS galaxies show a range of rest-frame UV slopes, with $-2.7 < \beta < -1.0$, with many measured $\beta$-slopes dominated by large errors ($\Delta \beta > 0.5$).
Reassuringly, our $\beta$ values derived from the stacked photometry follow the distribution of individual values.
We compare our results to the extrapolated relations from fainter sources derived at $z \simeq 7$ using~\emph{HST} data by~\citet{Bouwens14}, and the two recent~\emph{JWST} results by~\citet{Topping23} at $z = 7.3$ and by~\citet{Cullen23} at $z = 8$--$12$ (here we use the redshift evolution applied to the slope found~\citealp{Rogers14}).
These studies computed $\beta$ by fitting a power law the available photometry probing the rest-frame UV.
At $z \simeq 6.7$ we find good agreement with these relations in our fainter bin, however we see that our brighter stack has a significantly bluer $\beta$-slope than expected from the extrapolated colour-magnitude relations from previous studies at fainter magnitudes.
At $M_{\rm UV} < -22$ the offset bluewards from the colour-magnitude relations is between $\Delta \beta \simeq 0.3$--$0.7$ depending on the study.
Looking at the slightly higher redshift bin at $z \simeq 7.2$ we find an offset to bluer $\beta$ values by $\Delta \beta = 0.4$ in both the brighter and fainter stack.
Our results support a flattening, and potentially even a turn-over, of the colour-magnitude relationship at \muv$\lesssim-22$, with these galaxies showing a mean colour of $\beta = -2.1$ in contrast to the predicted colour of $\beta \simeq -1.4$ to $-1.7$ from the relations extrapolated from fainter LBGs.
As we discuss further in Section~\ref{sect:disc}, this turn over can be explained by the effect of scatter in the obscuration when considering sources that have a steeply declining number density. 

In the measurement of the colour-magnitude relationship we must consider any effect of sample selection and $\beta$-measurement bias in the results we obtain.
It is possible that we could be missing redder $z \simeq 7$ galaxies due to the requirement that the sources show good high-redshift fits and poorer quality fits to (typically redder) low-redshift galaxy contaminants.
As shown in Fig.~\ref{fig:cm} we are able to measure $\beta$-values as red as $\beta \simeq -1.2$ for the sources in our sample, even at the faint end, whereas we do not find significant numbers of the brightest sources to be as red (even though the increased S/N should make bright, red, sources easier to identify than similarly red, fainter sources).
The REBELS sample selection is not only based on the rest-frame UV bands but also includes the \chone~and \chtwo~bands in the SED fitting.
The~\emph{Spitzer}/IRAC colour aids in the selection of robust $z \simeq 7$ galaxies due to the specific colours produced by the rest-frame optical nebular emission lines in the \chone~and~\chtwo~bands~\citep{Smit15b}.
Using these filters could be biasing our sample towards bluer slopes by potentially selecting young galaxies with stronger nebular emission.
We discount this however, as the distribution of the ${\rm EW}_{0}({\rm H}\beta + {\rm [OIII]})$ of the REBELS sample is in excellent agreement with $z \simeq 7$ samples that are selected only based on a strong Lyman break (see figure 18 of~\citealp{Bouwens22}).
In fact, because these colours are challenging to reproduce by low-redshift galaxy contaminants it can aid in the recovery of good high-redshift galaxy fits to sources with redder rest-frame UV slopes (e.g. in~\citealp{Endsley21}; see Stefanon in prep. for individual SED fits).
Hence we conclude that our measurements of the rest-frame UV slope of the REBELS LBGs are unlikely to be significantly biased, with the caveat that we only select sources that are bright in the rest-frame UV, and hence will be incomplete to the most obscured galaxies (with the extreme situation being fully `UV-dark' galaxies as found in e.g.~\citealp{Fudamoto21}).

\subsection{$\mathbf{IRX}$--$\mathbf{\beta}$~relation at $\mathbf{z \simeq 7}$ from REBELS}

\begin{figure}

\includegraphics[width = 0.45\textwidth]{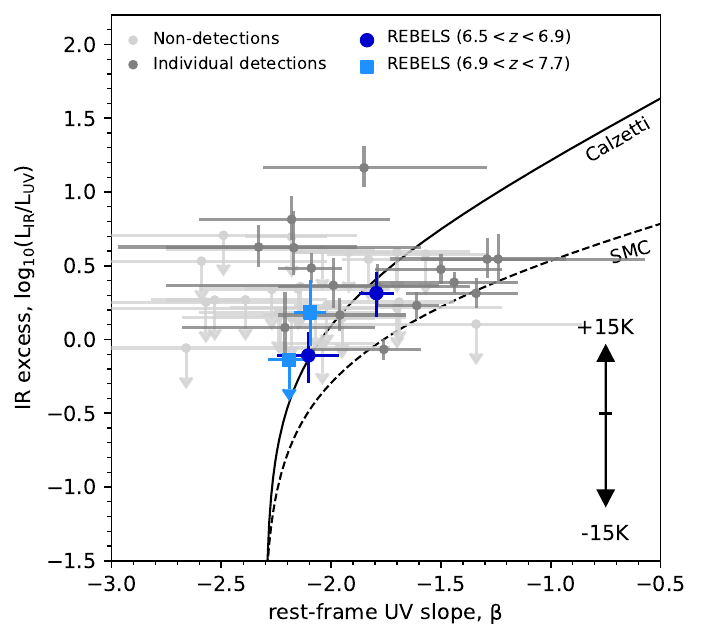}

\caption{The \irxb~relation as derived from the REBELS sample.
The individual dust continuum detected galaxies are shown as the filled dark grey points, with undetected galaxies shown as the lighter grey upper limits.
The results from our stacking analysis are shown as the dark blue circles (light blue squares) at $z \simeq 6.7$ ($z \simeq 7.2$).
Within each redshift bin the brighter \muv~stack is found to have a redder $\beta$ and a larger IRX.
We assume a dust temperature of $46\,{\rm K}$ and emissivity of $\beta_{\rm d} = 2.0$, with the arrow shown in the lower right of the plot illustrating the systematic uncertainty we would obtain assuming $\pm 15\,{\rm K}$.
The expected relation for Calzetti-like dust assuming $\beta_{0} = -2.3$ is shown as the black solid line, with the expected relation from an SMC extinction curve shown as the dashed line.
The right axis shows the dust attenuation in magnitudes corresponding to the IRX according to the Calzetti dust law with a screen geometry.
}\label{fig:irx}
\end{figure}

In Fig.~\ref{fig:irx} we present the \irxb~relation derived from the REBELS sample in the redshift range $6.5 < z < 7.7$.
We also show the derived values for individual sources, of which there are 18 detections at $ > 3.3\sigma$ from~\citet{Inami22}.
These results (both individual galaxies and for the stacks) were computed with the assumption of a modified blackbody FIR SED, with an assumed dust temperature of $T = 46\,{\rm K}$ and $\beta_{\rm d} = 2.0$ (Section~\ref{sect:fir}).
The individual results show a large scatter horizontally on the plot as a result of the large errors in individual measurements of the rest-frame UV slope.
We find that the majority of this range in observed rest-frame UV slopes in the sample can be explained with statistical scatter, with the intrinsic variation as a function of \muv~derived to be of the order of $\Delta \beta = 0.1$--$0.3$ for REBELS (see Table~\ref{table:flux}).
Rather the scatter can be explained simply due to the large errors on the individual $\beta$ measurements, which we have demonstrated via a simple simulation assuming the sample is drawn from a constant input $\beta = -2.0$ with the same $\beta$ measurement errors.
With our assumed rest-frame FIR SED, based on the work of~\citet{Sommovigo22a}, we do not confirm any sources significantly below the SMC-like \irxb~relation as found by previous high-redshift studies (e.g.~\citealt{Barisic17, Faisst17, Smit18}; note that these works assumed a lower $T_{\rm d} \simeq 30$--$40$).
Although at the depths of our observations, 31 of the 49 sources in REBELS are undetected in the dust continuum and hence the IRX values represent upper limits.
We see one source that is significantly in excess of the others, with ${\rm IRX} = 1.2 \pm 0.2$.
This is the unusually FIR bright object REBELS-25 that is discussed further in~\citet{Hygate23}.
This source is included in our stacks, however our results are unchanged if it is removed.
These data are shown in comparison to the expected relation for a~\citet{Calzetti00} dust attenuation and SMC dust extinction law, assuming an intrinsic $\beta$-slope, $\beta_{0} = -2.3$.
This intrinsic slope is consistent with that found in detailed SED fitting of comparable mass sources at $z = 3$~\citep{McLure18} and similar to that found in simulations of galaxies at $z \simeq 5$ (e.g.~\citealp{Cullen18} found $\beta_{0} = -2.4$).
We present a fit to the \irxb~relation, and fits from previous works at higher redshift, that include a steeper intrinsic $\beta$ in Section~\ref{sect:irxalpine}.
We find no strong correlation between the offset from a Calzetti-like relation and $A_{\rm V}$, \mstar, or spatial offset between the rest-frame UV and FIR flux (from~\citealt{Inami22}).
There is a weak trend that the FIR brightest galaxies tend to be above the relation, such as REBELS-25 (as has been seen for ULIRGS; see discussion below).

Turning to the stacked results, we present four individual points corresponding to the two different redshift and \muv~bins.
For the $6.5 < z < 6.9$ sub-sample we see that the brighter and fainter stack shows significantly different rest-frame UV slopes, with the brighter stack ($M_{\rm UV} < -22$) appearing bluer, while simultaneously being fainter in the FIR.
The brighter stack also shows a lower ALMA detected flux, and hence a lower IRX both from a higher \Luv~and a reduced \Lir.
The same trend is seen for the galaxies in the $6.9 < z < 7.7$ sub-sample, however here as both stacks are blue in the rest-frame UV (and the brighter stack is undetected in the FIR), we have a reduced dynamic range in $\beta$.
Overall, we find, somewhat counter-intuitively to the consensus colour-magnitude relation (Section~\ref{sect:cm} and Fig.~\ref{fig:cm}), that the rest-frame UV~\emph{brightest} galaxies in REBELS are bluer than the sources at slightly fainter magnitudes.
As expected by the canonical \irxb~relation, we find the bluer sources show a reduced IRX, and this is driven primarily due to a reduced \Lir~(although as we have previously described, note that galaxy age, dust SED and star-dust geometry can alter the expected relation; e.g.~\citealp{Popping17}).

\subsubsection{Comparison to previous studies at $z \simeq 7$}

\begin{figure}

\includegraphics[width = 0.45\textwidth]{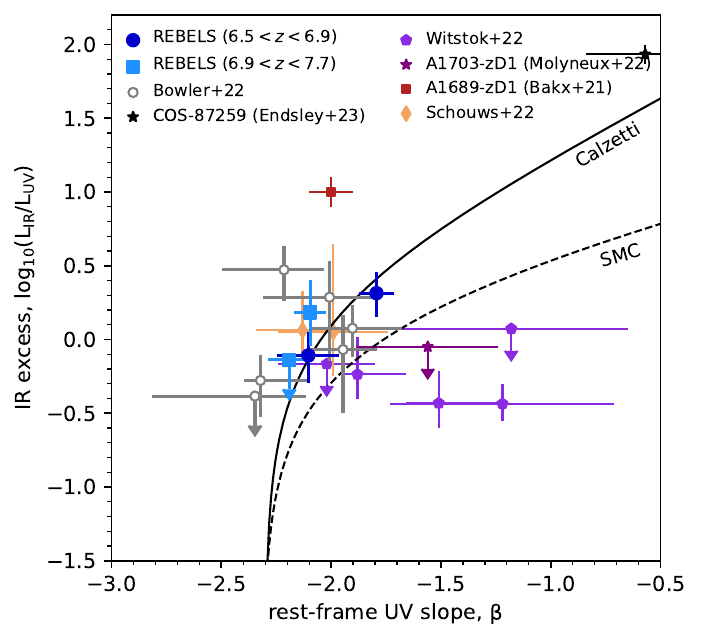}
\caption{The stacked \irxb~points from our analysis of the REBELS sample in comparison to previous results for LBGs at $z \simeq 7$.
We show the stacked results from~\citet{Schouws22} as the orange diamonds.
The six individual galaxy results of~\citet{Bowler22} are shown as the open grey circles.
Further sources from~\citet{Witstok22},~\citet{Bakx21} and~\citet{Molyneux22} are shown as the purple diamonds, red square and purple diamond respectively.
We show the radio-loud AGN identified initially as a bright $z \simeq 7$ LBG by~\citet{Endsley23} as the black star in the upper right.
}\label{fig:irxcomp}
\end{figure}

In Fig.~\ref{fig:irxcomp} we compare our REBELS results with those from previous studies at $z \simeq 7$.
We find that our stacked results are in good agreement with the previous measurements derived from luminous LBGs in~\citet{Schouws22} and~\citet{Bowler22}.
There are a handful of sources with redder rest-frame UV slopes and low IRX-values found within the study of~\citet{Witstok22}, although we note that the measured $\beta$-slopes are relatively uncertain in these cases.
\citet{Molyneux22} found a red rest-frame UV slope and an upper limit on the dust continuum emission in the $z = 6.8$ galaxy A1703-zD1.
These studies all assumed a dust temperature of $50\,{\rm K}$ and $\beta_{\rm d} = 1.5$--$1.6$ and hence we expect no appreciable offset to the results of this work due to differences in the chosen rest-frame FIR SED.
As shown in Fig.~\ref{fig:dustT} and further discussed in Section~\ref{sect:irxalpine}, although the dust temperature in these cases is higher than we assume in this work, the lower $\beta_{\rm d}$ compensates almost exactly.
We additionally show two galaxies where there is a confirmed ALMA detection and robust rest-frame UV slope determination.
The recent study of the $z = 7.13$ lensed galaxy A1689-zD1 from~\citet{Bakx21} found an IRX value that is in excess of he majority of the other points and the canonical Calzetti-like relation.
The \Lir~of this source was derived with the observed best-fitting dust temperature of $T_{\rm d} = 40\,{\rm K}$, with a fixed $\beta_{\rm d} = 2.03$.
If a higher dust temperature was assumed (to make the FIR analysis consistent with this work, and the other studies shown in this plot) this would increase the IRX by $0.2\,{\rm dex}$ (Fig.~\ref{fig:dustT}) resulting in an even greater excess.
To show this source on the~\irxb~relation we take the rest-frame UV colour derived by~\citet{Watson15}.
\citet{Knudsen17} argue that A1689-zD1 could be a massive starburst due to the observed \cii deficit, large~\Lir~(given the stellar mass) and disturbed morphology.
The fact that A1689-zD1, as well as REBELS-25 in Fig.~\ref{fig:irx}, appear in the upper left region of the~\irxb~diagram could be due to a spatial offset between the regions emitting in the rest-frame UV and FIR.
The FIR emission in this case would be dominated by optically thick emission, while the rest-frame UV colour is measured from unobscured stars leading to an unusually blue colour for the observed IRX (as seen in ULIRGS;~\citet{Casey14}, and predicted in theoretical works e.g.~\citealp{Popping17, Behrens18, Liang19, Sommovigo20, Ferrara22}).
We also show the galaxy COS-87259 from~\citet{Endsley23} that was found within the COSMOS field using an LBG selection, but has been confirmed to be a highly star forming and dust obscured radio-loud AGN at $z = 6.853$.
This source is very red, but it has a high derived IRX placing it slightly above the prediction of a Calzetti-like~\irxb~relation.

\subsubsection{Individual REBELS galaxies at $z > 7.7$}

\begin{figure}

\includegraphics[width = 0.45\textwidth]{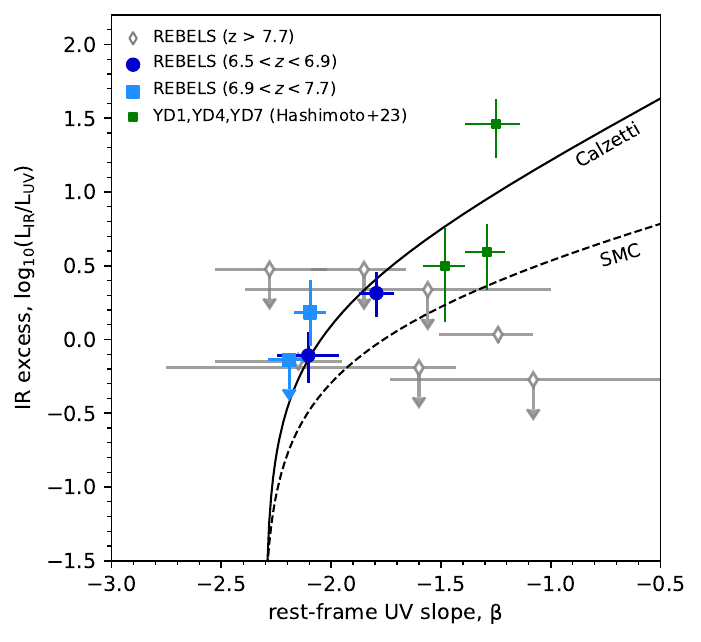}
\caption{The individual \irxb~points for the seven REBELS galaxies with photometric redshifts at $z > 7.7$ (open grey diamonds) in comparison to our stacked results at $z = 6.5$--$6.9$ (navy circles) and $z = 6.9$--$7.7$ (blue squares).
We also compare to the $z = 7.88$ galaxy group found within the Abell cluster A2744 lensing field, that have been detected in the dust continuum by~\citet{Hashimoto23}.
The Calzetti and SMC-like relations are shown as described in the caption of Fig.~\ref{fig:irx}.
}\label{fig:irxhighz}
\end{figure}

In Fig.~\ref{fig:irxhighz} we show the \irxb~results we derive for the seven galaxies in REBELS that have photometric redshifts at $z > 7.7$.
These galaxies were not included in our stacking analysis due to four sources having Band 7 observations, and the relatively uncertain photometric redshifts derived for these sources at the very high-redshift end of the REBELS sample.
We compare to the~\citet{Hashimoto23} work that spectroscopically confirmed a group of galaxies (nicknamed RIOJA) at $z = 7.88$, with three of the components showing detections in the rest-frame FIR from ALMA.
Other $z \gtrsim 7.5$ sources have been observed with ALMA (e.g. MACS0416\_Y1 at $z = 8.31$ and MACS0416-JD at $z = 9.11$;~\citealp{Hashimoto18, Bakx20}) however these other objects do not have published rest-frame UV slopes.
Two of the REBELS sources at $z > 7.7$ are detected in the dust continuum (REBELS-4 and REBELS-37; also called XMM-355 and UVISTA-1212 respectively in~\citealp{Bowler20}), while the other five are not.
REBELS-4 is in good agreement with our stacked results at slightly lower redshift, however REBELS-37 shows a redder rest-frame UV colour and deficit in IRX from both the Calzetti and SMC relations shown.
We note that the $\beta$-slope value is more uncertain at these redshifts, due to the few bands ($H, K_s$) available for fitting, and the broader uncertainty in photometric redshift leading to a degeneracy between redshift and slope.
Excluding REBELS-37, we find good agreement within the (large) errors with our $z \simeq 7$ stacks, although we note that the majority are upper limits on the dust continuum.

\subsection{$\mathbf{IRX}$--$\mathbf{\beta}$~from $\mathbf{z = 4-8}$ from ALPINE and REBELS}\label{sect:irxalpine}

To provide a consistent comparison to the $z > 6.5$ results from REBELS, we performed the same analysis on the ALPINE sample in the COSMOS field (see Section~\ref{sect:data}).
We present the ALPINE results compared to the individual object measurements in Fig.~\ref{fig:irxalpine} with the values presented in Table~\ref{tab:alpine_flux}.
As the ALPINE sample is both larger and has a broader range of measured stellar masses, we additionally binned in \mstar.
In our computation of the \Lir~and hence IRX for the two samples we assumed the same rest-frame FIR SED, with a dust temperature of $T_{\rm d} = 46\,{\rm K}$ and emissivity of $\beta_{\rm d} = 2.0$ following the work of~\citet{Sommovigo22a, Sommovigo22} (see Section~\ref{sect:fir}).
We compare the ALPINE results with those from REBELS in Fig.~\ref{fig:irxrebelsalpine}.
As was found for the REBELS sources, the brighter rest-frame UV stacks have lower IRX and appear bluer.
We also see a strong dependence on stellar mass, with the galaxies at \lmstar $> 10$ showing considerably redder colours with $\beta \gtrsim -1.75$ and a higher IRX by $\simeq 0.75$.
We note here that due to the selection methodology of the initial ALPINE sample, a larger fraction of the sources in the $z \simeq 5.5$ bin were selected to be Lyman-$\alpha$ emitters.
Taking a rest-frame EW of $> 50 (25)\,$\AA~as the separation between LBGs and LAEs, 8 (30) percent of the $z = 4.5$ sub-sample are LAEs in comparison to 38 (57) percent of the $z = 5.5$ sub-sample.
As LAEs have in general been found to show lower dust attenuation (e.g.~\citealp{Schaerer15}) this could explain the small offset we see between these two redshift bins.

As we discuss further in the next section, we find consistent results between the derived~\irxb~relation between the REBELS and ALPINE samples in our analysis, when we use the same modified blackbody SED fitting analysis and $\beta$ measurement procedure.
The data appears to agree with the local starburst relation of~\citet{Calzetti00}, with no evidence from our stacked results for a deficit in the relation that could be consistent with SMC-like dust, given our assumptions on FIR SED.

By combining the results from the two surveys we are able to measure the \irxb~relationship across a wide redshift range from $z = 4$--$8$ from the largest sample of \lmstar$> 9$ galaxies available with deep ALMA follow-up.
Taking the stacked detections for ALPINE and REBELS, we fit the slope of the \irxb~relation with a given intrinsic rest-frame UV slope, $\beta_0$, according to the formalism presented in~\citet{McLure18} as:

\begin{equation}
    {\rm IRX} = 1.71\times 10^{(0.4\,{\rm d}A_{1600}/{\rm d}\beta\,(\beta - \beta_{0}))} - 1)
\end{equation}

In this formalism, the Calzetti (SMC)-like relation has a slope of ${\rm d}A_{1600}/{\rm d}\beta = 1.97 (0.91)$ and the 1.71 pre-factor is a constant set by the bolometric correction between the total rest-frame UV emission available to heat the dust and that characterised by \Luv.
The pre-factor can change if we break the assumption of a dust screen, however for this analysis we keep it constant.
For our combined ALPINE and REBELS results we find a best-fitting slope of ${\rm d}A_{1600}/{\rm d}\beta = 2.11 \pm 0.13$ when assuming $\beta_0 = -2.3$ or a shallower slope of the \irxb~relation of ${\rm d}A_{1600}/{\rm d}\beta = 1.38 \pm 0.09$ when assuming $\beta_0 = -2.5$.
The intrinsic rest-frame UV slope of our sample is not known, however from {\sc BAGPIPES} SED fitting analysis we find it to be between $\beta_{0} = -2.3 $ and $-2.5$ and hence present the results of both fits.
As can be seen in Fig.~\ref{fig:irxrebelsalpine}, both $\beta_{0}$ assumptions provide a good description of the data over a broad range in measured rest-frame UV slope.
Our results are in general in excess of the previously derived \irxb~relations at $z > 4$ (e.g. from~\citealp{Fudamoto20, Schouws22}).

\subsubsection{Comparison to previous results from the ALPINE survey}
Our conclusions on the \irxb~relation at $z = 4$--$8$ are different to those found in the previous ALPINE analysis presented in~\citet{Fudamoto20}, particularly at \lmstar$> 10$ where we find a higher IRX by around $0.5\,{\rm dex}$ when comparing stacks across the same \mstar~range.
The later studies of~\citet{Burgarella22} and~\citet{Boquien22} found similar conclusions to the~\citet{Fudamoto20} study with further analysis of subsets of the ALPINE sample.
\citet{Fudamoto20} present stacked \irxb~relations using bins in $\beta$ and \mstar, finding the results to be consistent.
In the following discussion we compare to the \mstar~binning results of~\citet{Fudamoto20} as this has been shown to be the least biased estimator of \irxb~ (e.g.~\citealp{McLure18}).
This provides the most natural comparison as our points are already stacked in \mstar, however we additionally stack in \muv~bins.
Hence in the following discussion we combined our \muv~bins at a given \mstar.
This leads to points that lie mid-way between the two \muv~bins per \mstar bin, as expected.
To identify the cause of this offset we first directly compared the derived $\beta$-slopes, \muv~values and ALMA fluxes for individual objects.
We find that our rest-frame UV slopes are on average bluer than those derived in ALPINE by $\Delta \beta = 0.1$, with around half of this difference attributed to the dust law that we assume in the SED fitting (we assume Calzetti, whereas in~\citealp{Faisst20} took the average between results with an SMC and Calzetti dust law).
The \muv~values are found to be offset slightly brighter (0.1 mag) in our analysis, which used the COSMOS2020 catalogue instead of the COSMOS2015 data analysed in~\citet{Faisst20}, however this has a negligible effect on the derived IRX.
For the 20 percent of the ALPINE sample that have dust continuum detections we find good agreement between our raw flux measurements.
Both our study and that of~\citet{Fudamoto20} take into account the fact that the dust continuum emission may be extended in the higher mass (\lmstar$> 10$) stacks by using a Gaussian fit to the ALMA data.
On closer inspection, the extension found in these stacked images is due to both an intrinsic extension (i.e. higher mass sources have an extended dust distribution) and an artificial extension introduced in the stacking process due to offsets between the rest-frame UV and FIR centroid.
In the binning analysis, we determine the $\beta$ slopes from SED fitting to the stacked optical/NIR photometry, while~\citealp{Fudamoto20} take the median $\beta$ in each bin.
However despite the different method, when comparing the same bins in \mstar~we find only a $0.1$ difference between the resulting $\beta$-slopes ($0.2$ for the lower mass bin at $z \simeq 4.5$), and $0.05$ of the difference can be accounted for by the different assumed dust law in the fitting (Section~\ref{sect:methods}).

\begin{figure}

\includegraphics[width = 0.45\textwidth]{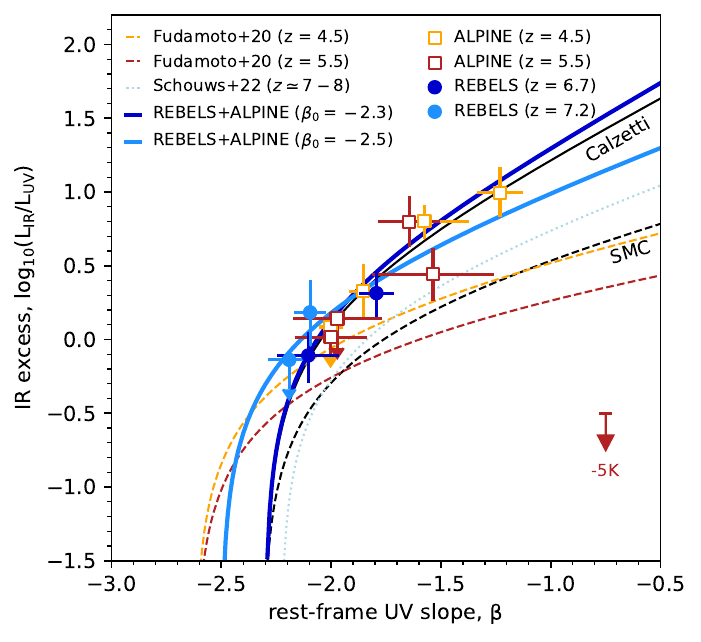}

\caption{ The~\irxb~relation derived in this study from the ALPINE and REBELS surveys at $z \simeq 4$--$8$.
The ALPINE points at $z = 4.5$ ($z = 5.5$) are shown as the orange (red) open squares, while the REBELS points at $z = 6.7$ ($z = 7.3$) are shown as the blue (light blue) circles. 
Our best fitting \irxb~relation to these data are shown as the solid blue (light blue) lines for an assumed intrinsic rest-frame UV slope of $\beta_{0} = -2.3 (-2.5)$.
We also show the best-fitting~\irxb~relation from~\citet{Schouws22} (assuming a $\beta_{0} = -2.23$) as the blue dotted line.
The best-fitting~\irxb~relations assuming a $\beta_{0} = -2.62$ from~\citet{Fudamoto20} are shown as the orange (red) dashed lines for $z = 4.5$ ($z = 5.5$).
The difference between our ALPINE results and those of~\citet{Fudamoto20} can be mostly explained by the different assumed FIR SED.
\citet{Fudamoto20} assumed $T_{\rm d} = 41\,{\rm K}$, in comparison to the $T_{\rm d} = 46\,{\rm K}$ assumed in this work, leading to the difference illustrated in the bottom right as the red arrow.
}\label{fig:irxrebelsalpine}
\end{figure}

\begin{figure}
\includegraphics[width=0.49\textwidth]{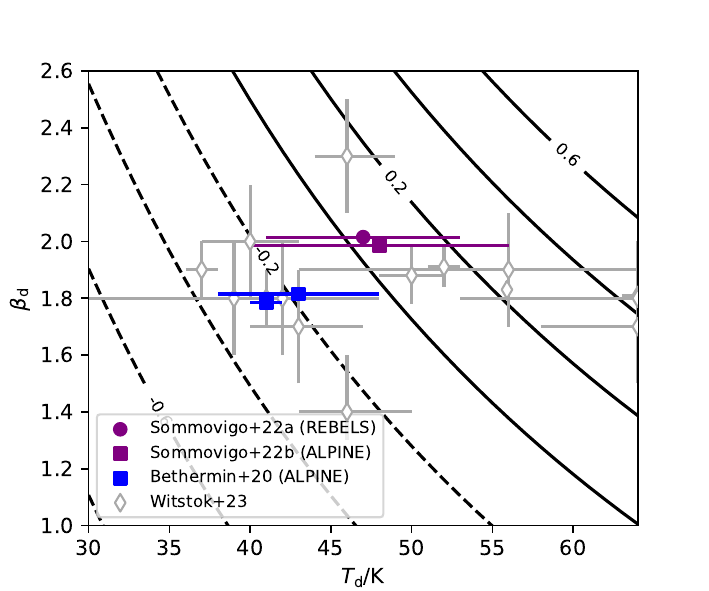}
\caption{The modified blackbody fitting parameters from previous observations and models.
We show the dust temperature against the emissivity coefficient, for an optically thin model.
The results for the theoretical analysis by~\citet{Sommovigo22a, Sommovigo22} are shown as the purple circle (square) for REBELS (ALPINE).
In the ALPINE survey,~\citet{Bethermin20} measured an empirical SED that had equivalent modified blackbody parameters as shown by the blue squares (the $z = 4$ point has the smaller error bar, with the $z = 5$ point showing a higher temperature).
For both these studies the $\beta_{\rm d}$ value was fixed.
The contours show the offset computed in the derived IRX value from the parameters used in this work ($T_{\rm dust}= 46\,{\rm K}$, $\beta_{\rm d} = 2.0$).
Such that the~\citet{Bethermin20} parameterisation would give a lower IRX by $\sim 0.25\,{\rm dex}$.
The results of fitting with a self-consistent (not optically thin) model to $z = 4$--$8$ galaxies with multiple FIR data points in~\citet{Witstok23} are shown as the grey open diamonds, demonstrating the current uncertainties and potential intrinsic scatter on these parameters at high redshift.
}\label{fig:dustT}
\end{figure}

Assuming that the fluxes in the stacks are consistent, the main difference is in the FIR SED assumed in the derivation of the \Lir.
\citet{Fudamoto20} used a scaling factor to compute \Lir~that was derived from an empirical FIR template created by stacking~\emph{Herschel} data in the COSMOS field.
An expanded sample of photometrically selected galaxies over a similar redshift to the ALPINE sample was used in the creation of this template~\citep{Bethermin20}, and it can be approximated with a modified blackbody with a fixed $\beta_{\rm d} = 1.8$ of temperature $T_{\rm d} = 41\,\pm 1\,{\rm K}$ and $T_{\rm d} = 43\,\pm 5\,{\rm K}$ at $z = 4$--$5$ and $z = 5$--$6$ respectively.
While only $3$--$5\,{\rm K}$ lower than that assumed in this work, the difference is enough to account for a $0.25\,{\rm dex}$ difference in the resulting IRX given the same input flux measurement.
We mark this offset as an arrow in Fig.~\ref{fig:irxrebelsalpine}.
This is illustrated in Fig.~\ref{fig:dustT}, where we show the offset in IRX expected for changes in $T_{\rm d}$ and $\beta_{\rm d}$.
This difference, and the slightly bluer rest-frame UV slopes we find, can account for $0.3\,{\rm dex}$ of the observed difference between our analysis of ALPINE and that presented previously in~\citet{Fudamoto20}.
As can be seen in Fig.~\ref{fig:irxalpine}, the ALPINE sample consisted of 80 percent upper limits on the dust continuum emission and hence the exact stacking process could contribute to the remaining difference.
Despite not knowing exactly the dust SED for the ALPINE and REBELS samples, we have shown that the two samples have consistent IRX relations when the ALMA measurements are fit with the same modified blackbody assumptions.
If the ALPINE sample does indeed show a lower dust temperature to that of REBELS (as expected from the $T_{\rm d}$-redshift relation in e.g.~\citealp{Schreiber18}), then we would recover a lower \irxb~relation for the ALPINE dataset by $0.25\,{\rm dex}$.

\section{Discussion}\label{sect:disc}

This work presents a consistent analysis of the two most substantial ALMA surveys measuring the dust continuum emission from $ z > 4$ LBGs.
We have computed the~\irxb~relation at $z \simeq 7$ from REBELS, and compare this to a consistent analysis of the $z = 4.5$ and $z = 5.5$ samples from ALPINE.
These surveys targeted galaxies with stellar masses of \lmstar $> 9$ and hence probe the high-mass end of the known galaxy distribution at these redshifts.
Even before considering the results from the ALMA observations themselves, we find that the REBELS galaxies are surprisingly blue in the rest-frame UV as compared to the predicted colour of galaxies from the extrapolated colour-magnitude relation found for fainter sources (Fig.~\ref{fig:cm}).
Despite being some of the most massive galaxies known at $z \simeq 7$, the REBELS sources show a flatting, or a potential turnover, in the colour-magnitude relation for galaxies at \muv$<-21.5$.
As discussed in Section~\ref{sect:cm}, we do not think this behaviour is due to our sample selection in part due to the fact that predominantly blue $\beta$-slopes are found even in the most luminous (and hence highest S/N) galaxies where redder slopes would be robustly measured.
Following the arguments presented in e.g.~\citealp{Bowler15, Bowler20} amongst other studies (e.g.~\citealp{Salim12}), when considering galaxies found at the bright-end of the rest-frame UV luminosity function (UVLF), the effect of scatter must be considered.
If we consider an underlying stellar mass function (SMF) with an exponential cut-off above the characteristic mass, and estimate from this the expected UVLF, given the scatter between \mstar~and the \Luv, we expect to see a shallower decline in the number density to brighter galaxies.
Because of the steepness of the SMF, galaxies that are found to be bright in the rest-frame UV are necessarily the sources that have low dust attenuation.
Galaxies of the same stellar mass, but with a higher than average attenuation would instead be scattered fainter on the UVLF and be lost within the large population of fainter sources.
The majority of the REBELS sources were selected from wide-area ground-based data and they represent a very rare population of galaxies that sit brightward of the knee of the UVLF.
Hence from the arguments detailed above, we would expect them to show a relatively low dust obscuration given their stellar mass (e.g. as found in~\citealp{Algera23}), and hence a lower IRX and bluer $\beta$.
An additional factor that could contribute to the unexpectedly blue colours we observe for the REBELS sources is a geometric offset between the dust and stars within the galaxy (e.g. as predicted by~\citealp{Popping17}).
The UV variability model of~\citet{Shen23} predicts that the rest-frame UV brightest galaxies should be blue, due to their young ages and clearance of dust in the star-burst phase.
A clumpy morphology with an offset between the observed young stars stars and dust, leading to relatively unobscured regions, has already been observed in higher-resolution ALMA observations of REBELS galaxies in~\citet{Bowler22} supporting this hypothesis.

\subsection{Implications for the dust attenuation curve at $\mathbf{z > 4}$}
We find that the~\irxb~relation derived from a stacking analysis of the ALPINE and REBELS samples are consistent with a Calzetti-like relationship as found at $z = 0$, with our assumed rest-frame FIR SED for both samples.
We find no evidence that the sources~\emph{on average} lie below the relation, which could indicate a different attenuation law such as an SMC-like relation (discounting the effect of geometry which tends to make the observed~\emph{attenuation} law appear shallower even in the case of a steeper~\emph{extinction} law).
As shown in Fig~\ref{fig:irxrebelsalpine}, our results differ from those found using the ALPINE dataset by~\citet{Fudamoto20}, who found a deficit in IRX for a given rest-frame UV slope and concluded that an SMC-like attenuation curve was preferred.
As discussed in Section~\ref{sect:results}, the difference between our results and those of~\citet{Fudamoto20} is primarily due to the assumed rest-frame FIR SED (with this work using a higher dust temperature by $5\,{\rm K}$), with a minor effect of our analysis deriving bluer rest-frame UV slopes from the stacked photometry ($\Delta \beta = 0.1$).
The best-fit relation presented in~\citet{Fudamoto20} assumed an intrinsic $\beta_{0} = -2.62$ (fits were also presented for redder $\beta_{0}$) and showed a tentative evolution in the normalisation, such that the IRX is decreasing at a given $\beta$-slope from $z = 4.5$ to $z = 5.5$.
We instead find little evolution in the~\irxb~relation from $z \simeq 4$ to $z \simeq 7.2$ from our analysis.
This relies on our assumption that the FIR SED (and intrinsic rest-frame UV slope) remains approximately constant between the samples and redshift ranges, as found by~\citet{Sommovigo22a, Sommovigo22} which motivated the $T_{\rm d}$ and emissivity coefficient used in this work for the ALPINE and REBELS datasets.
If instead the dust temperature differs between the two samples at above and below $z = 6$, then we would conclude that the REBELS sources lie above the ALPINE galaxies on the~\irxb~plane by approximately $0.3\,{\rm dex}$.
This would potentially indicate a different selection function between the surveys (e.g. we already know that ALPINE targeted more Lyman-$\alpha$ emitters than REBELS; see Section~\ref{sect:caveats}).

Our results are also higher (particularly at redder $\beta$) than the best-fit derived by~\citet{Schouws22} (assuming an intrinsic $\beta_{0} = -2.23$, at $T_{\rm d} = 50\,{\rm K}$).
However the stacked results from that work are consistent within the errors with our findings from REBELS (see Fig.~\ref{fig:irxcomp}).
In the range of stellar mass between \lmstar$= 9$--$10$, galaxies in both REBELS and ALPINE appear blue ($\beta \simeq -2$) and the obscured fraction is $0.4$--$0.6$ (Table~\ref{table:sed}).
For the higher mass galaxies found within ALPINE (as REBELS contains very few galaxies at \lmstar$> 10$ assuming a parametric SFH; although see~\citealp{Topping22}) we find redder slopes ($\beta \simeq -1.5$) and obscured fractions approaching 0.9.
Comparing our results to studies that targeted lower mass galaxies there is some evidence for a difference in IRX with galaxy luminosity (or stellar mass).
For example the stacking analysis of~\citet{Bouwens16} and~\citet{Bouwens20} derived from the deep ALMA data covering the~\emph{Hubble} Ultra Deep Field showed only upper limits over the broad redshift range $z = 4$--$10$.
\citet{Bouwens20} also found evidence for an SMC-like attenuation curve at \lmstar$ < 9.25$, while a Calzetti like curve was preferred at higher masses.
These results are consistent with our analysis, which supports a Calzetti-like attenuation curve for \lmstar$> 9$ from $z = 4$--$8$.
Our find that there is a lack of evolution in the attenuation curve from local starburst galaxies up to $z \gtrsim 7$ could suggest that the conditions required to form Calzetti-like dust are present already 800 Myr after the Big Bang.
As shown by various theoretical studies however, taking the observed~\irxb~relation as evidence for a particular dust attenuation curve can be problematic due to the myriad of other factors that can influence the observed relation (e.g.~\citealp{Popping17, Narayanan18}).
The fact that we do not see a significant excess above the Calzetti relation further suggests that on average the REBELS and ALPINE galaxies are not dominated by significant optically thick regions contributing to the \Lir, that would boost the measured IRX values (e.g. as seen in the case of ULIRGs;~\citealp{Casey14}).

\subsubsection{Evidence for a high gas-phase metallicity?}
Further insight into the origin of our results for massive galaxies can be gained from the works of~\citet{Shivaei20a, Shivaei20} who found that the dust attenuation curve correlated most strongly with gas-phase metallicity from a detailed study of sources at $z = 2$--$2.5$. 
They found that at $12 + {\rm log}({\rm O/H}) > 8.5$ a Calzetti-like attenuation curve was consistent with the data, with a steeper attenuation curve found for lower metallicity sources.
Qualitatively our results agree with these $z \simeq 2$ studies, although the difference in methodology makes it non trivial to compare quantitatively.
In particular,~\citet{Shivaei20} remove sources with very strong emission line strengths (\oiiio $> 630$\AA) that we inferred to be present in the majority of the REBELS sources~\citep{Bouwens22}.
In addition, they compute the IR luminosity using a template from~\citet{Rieke09} for the more massive sources in their sample.
These templates are stated to be equivalent to a greybody curve with $T_{\rm d} = 38$--$64\,{\rm K}$ and $\beta_{\rm d} = 0.7$--$1$, which as shown in Fig.~\ref{fig:dustT} would give lower values of the IRX by $0.3$--$0.4\,{\rm dex}$.
With these caveats in mind, these previous results would suggest that the ALPINE and REBELS galaxies with \lmstar$\simeq 9.5$ show an increased gas-phase metallicity compared to lower mass sources, that then leads to the observed Calzetti-like dust attenuation curve.
If we take the mass-metallicity relation derived by~\citet{Sanders21} at $z \simeq 3$, we would predict a metallicity for this sample of $12 + {\rm log}({\rm O/H}) \simeq 8.3$.
We also obtain a comparable estimate of the metallicity using the fundamental mass-metallicity relation (FMR) from~\citet{Curti20} assumming a total SFR $\sim 50\,{\rm M}_{\odot}/{\rm yr}$ (Table~\ref{table:sed}). 
Recent results from early~\emph{JWST} analysis for lower mass galaxies have revealed similar FMRs as at $z \simeq 2$~\citep{Nakajima23} as well as evidence for a deficit from the FMR in galaxies at $z \simeq 7$~\citep{Curti23}.
If this deficit is found to hold for galaxies at \lmstar$\gtrsim 9.5$, this would imply a lower metallicity by $\sim 0.3\,{\rm dex}$ for galaxies in our sample given their \mstar.
Regardless of which of these metallicity calibrations we consider, the derived values of average metallicity quantitatively disagree with the results of~\citet{Shivaei20} that would predict that these sources should show an SMC-like dust attenuation curve at $12 + {\rm log}({\rm  O/H}) < 8.5$.
Despite this, our results are qualitatively in agreement with a picture where the ALPINE and REBELS sources have higher metallicities than lower-mass/fainter sources at the same redshifts, and hence a Calzetti-like dust attenuation law remains the preferred fit even up to $z \simeq 7$.
With upcoming NIRSpec observations we will be able to directly determine the extension of the FMR relation at $z = 7$ to higher masses, and test whether the sources that show significant dust detections are metal rich in comparison to lower mass (\lmstar$< 9$) galaxies.

\subsection{The IRX--$\mathbf{M_{\star}}$ relation at $\mathbf{z = 4}$-$\mathbf{8}$}

\begin{figure}

\includegraphics[width = 0.45\textwidth]{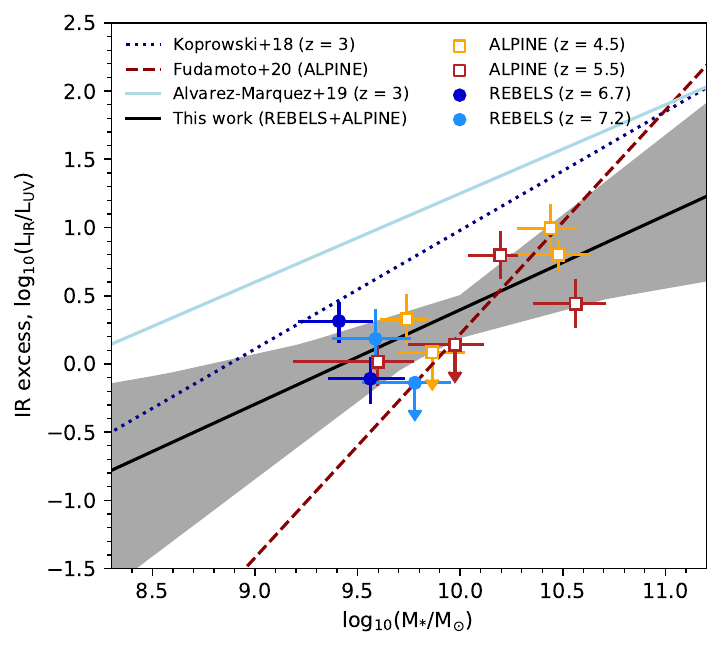}

\caption{The derived \irxm~values from our consistent analysis of the REBELS and ALPINE datasets.
The results from our REBELS analysis are shown as the dark blue (light blue) circles at $z \simeq 6.7$ ($z \simeq 7.2$).
The ALPINE results are shown as orange (red) open squares for the $z \simeq 4.5$ ($z \simeq 5.5$) sub-samples.
The best-fitting relation to these data is shown as the black solid line with the grey shading indicating the $1\sigma$ confidence regime.
Relations from previous studies at $z = 3$ from~\citet{Koprowski18} and~\citet{Alvarez-Marquez19} are shown as the dotted blue and solid light blue lines respectively.
The $z \simeq 4$--$6$ relation found by~\citet{Fudamoto20} is shown as the dashed red line.
}\label{fig:irxm}
\end{figure}

In principle the~\irxm~relation provides a more fundamental physical relationship than the~\irxb~plane, where the latter can be affected by large errors on the rest-frame UV slope measurement and geometric effects (e.g. see~\citealp{Faisst17, Liang19, Sommovigo20, Ferrara22}).
The stellar mass represents an integral of all past star-formation activity in the galaxy, and hence is expected at least qualitatively to be correlated with the production of dust (e.g. via stellar dust production and the overall metal enrichment of the ISM;~\citealp{Dayal22}).
\citet{Dunlop17} showed that stellar mass is a strong predictor of a FIR detection (and thus \Lir; see also~\citealp{McLure18, Bouwens20}).
Up to $z \simeq 3$ there has been in general a good agreement between previous studies of the \irxm~relation, with the results found to be approximately consistent to the local relation (e.g.~\citealp{McLure18, Koprowski18, Alvarez-Marquez19}; although see~\citealp{Reddy18}).
At higher redshift,~\citet{Fudamoto20a} found a steeper slope of the~\irxm~relation at $z = 2.5$--$4$ implying a reduced obscured SFR fraction and total SFR than the relations recovered at lower redshifts (at \lmstar $<11$ when the relations cross), a trend that was also recovered in the ALPINE sample~\citep{Fudamoto20}.
In Fig.~\ref{fig:irxm} we show the measured~\irxm~points from our stacking analysis of the ALPINE and REBELS samples.
We find an offset in the measurements from the $z = 3$ results of~\citet{Koprowski18} and~\citet{Alvarez-Marquez19}.
These studies use empirical templates in their derivation of the FIR luminosity, with the best-fitting templates used in~\citet{Koprowski18} showing a temperature of $40\,{\rm K}$.
Given the observed relation of $T_{\rm d}$ with redshift, and the lack of information on the FIR SED at $z > 6$ and observed changes in the FIR SED with physical properties (e.g. \mstar, SFR;~\citealp{Alvarez-Marquez19}), it is not trivial to compare IRX values over the redshift range $z = 2$--$8$ and be fully confident of the exact offsets between studies.
Nevertheless, we recover an offset of between $0.5$--$1.0$~dex when comparing our $z > 4$ results to those at $z \simeq 3$, depending on the lower redshift relation assumed.
Fitting these data points we derive the following~\irxm~relation:

\begin{equation}
    {\rm log}_{10}({\rm IRX}) = 0.69(\pm 0.12)\,{\rm log}_{10}\left (\frac{M_{\star}}{10^{10}\,{\rm M}_{\odot}}\right ) +0.40\,(\pm 0.05).
\end{equation}

While the derived slope is only weakly constrained due to the small dynamic range in stellar mass that we probe with the ALPINE and REBELS samples, it is consistent with the $z \simeq 3$ relations.
Further analysis of galaxies at \lmstar$<9$ will be required to confirm if the slope does steepen at $z \gtrsim 5$.
Due to the strong \mstar~dependence of obscuration however, these measurements are extremely challenging and must rely on stacking (for example the ASPECS program only found 18 ALMA detections from a sample of 1362 galaxies at $z = 1.5$--$10$, with the highest redshift being at $z = 3.7$;~\citealp{Bouwens20}).
Regardless of the exact form of the relation, the offset we observe in the derived IRX for a given \mstar~shows that between $z \simeq 2$ and $z = 4$--$8$ the degree of obscured star formation has dropped by a factor of $\sim 3$--$10$.
This is despite the sources showing good agreement with the galaxy ``main-sequence''~\citep{Topping22, Algera23} and hence they do not show a deficit in total SFR.
This finding is consistent with previous results that have derived a lower obscured fraction as a function of \mstar~at $z \gtrsim 5$~\citep{Fudamoto20, Gruppioni20, Algera23}, however we do not recover the steepness of these relations.
We note that recent results on the dust obscuration as a function of stellar mass from~\citet{Shapley23} using the Balmer decrement measured with~\emph{JWST} have not found evidence for evolution in the degree of dust obscuration between $z \simeq 2$--$6$, in contrast to our results.
One  explanation for these differing conclusions is that there is an evolution in the relation between continuum (as we measure in this work using the rest-frame UV slope) and the nebular reddening (as measured by~\citealp{Shapley23}).
Future rest-frame FIR observations of samples of massive $z > 7$ galaxies, such as those selected via upcoming wide-area NIR imaging observations from~\emph{Euclid} and~\emph{Roman}, will be required to confirm if the reduction in dust obscured star-formation continues to even higher redshifts.

\subsection{Caveats}\label{sect:caveats}
A key assumption in this work is that of a fixed dust SED with a dust temperature of $T_{\rm d}= 46\,{\rm K}$ and $\beta_{\rm d} = 2.0$.
These parameters were utilized within the REBELS survey as the fiducial rest-frame FIR SED and were derived by~\citet{Sommovigo22a}.
The dust temperature at $z > 5$ remains uncertain, and recent results have suggested that it may vary between galaxies from $T_{\rm d} = 35$--$90\,{\rm K}$ (e.g.~\citealt{Hashimoto19, Bakx20, Algera23a}).
For the emissivity index,~\citet{Witstok23} found a best-fitting value of $\beta_{\rm d} = 1.8 \pm 0.3$, consistent with our assumed model.
These results would suggest that a different dust temperature would be required for each source.
However, given the large errors for individual sources, and the lack of these measurements for the full sample, this analysis is unfeasible at this time.

Another important caveat is that the ALPINE and REBELS samples were selected in different ways, and for example the $z \simeq 5.5$ sample from ALPINE includes a larger fraction of Lyman-$\alpha$ emitters than REBELS.
The REBELS survey has no confirmed Lyman-$\alpha$ emission stronger than $EW_{0} = 25\,$\AA~\citep{Endsley22}, while in the $z = 5.5$ bin of ALPINE 57 percent of the sources have $EW_{0} > 25$\AA.
This comparison is complicated by the impact of the neutral IGM as we approach $z \simeq 7$, however as galaxies selected as Lyman-$\alpha$ emitters have been shown to have lower dust content than LBGs (e.g.~\citealp{Schaerer15, Matthee17}) this difference could cause a bias to lower IRX values in the ALPINE sample.
There is also evidence at lower redshifts that the dust attenuation curve varies between galaxies of the same stellar mass (e.g.~\citealp{Salim19}; see the review by~\citealp{Salim20}).
Using a subset of the ALPINE sample,~\citet{Boquien22} derived a range of best-fitting attenuation curves also suggesting that the attenuation law varies significantly between individual galaxies.
\citet{Narayanan18} have predicted using a simulation that the variation in attenuation curves between galaxies decreases with increasing redshift, however further observations are needed to understand the properties of dust within high-redshift galaxies.

Another factor to consider, with reference to our inferences about stellar mass dependence, is that we assumed a certain parametric star-formation history in computing these parameters.
\citet{Topping22} has shown that these measurement may underestimate the mass by on average $0.5\,{\rm dex}$.
This would cause an increased deficit (up to around $0.35\,{\rm dex}$, given the best-fit relation) in the~\irxm~relation to previous studies, however care must be taken to compare similar methodologies in determining \mstar.
Future observations of the REBELS galaxies with~\emph{JWST} (e.g. through program PID1626, PI Stefanon) will constrain the SFH and hence stellar masses with greater accuracy.

Finally, we consider the potential effects of sample incompleteness.
Due to the rest-frame UV based Lyman-break selection (or Lyman-$\alpha$ selection for some ALPINE sources), our samples will be incomplete to any highly obscured sources.
Such sources have been identified with ALMA, for example the `UV-dark' galaxies found serendipitiously via their \cii emission only $40$--$60\,{\rm pkpc}$ from the central (rest-UV bright) REBELS source in~\citet{Fudamoto21}.
It is challenging to place these types of galaxies on the \irxb~or~\irxm~relation as their rest-frame UV and optical are undetected, however it is likely that they are extremely red and massive (\lmstar $\lesssim 10$--$10.5$) and with a high IRX.
Further study is needed to understand this population of extremely obscured galaxies at $z \simeq 7$.

\section{Comparison to models}\label{sect:sim}
\begin{figure}

\includegraphics[width = 0.45\textwidth]{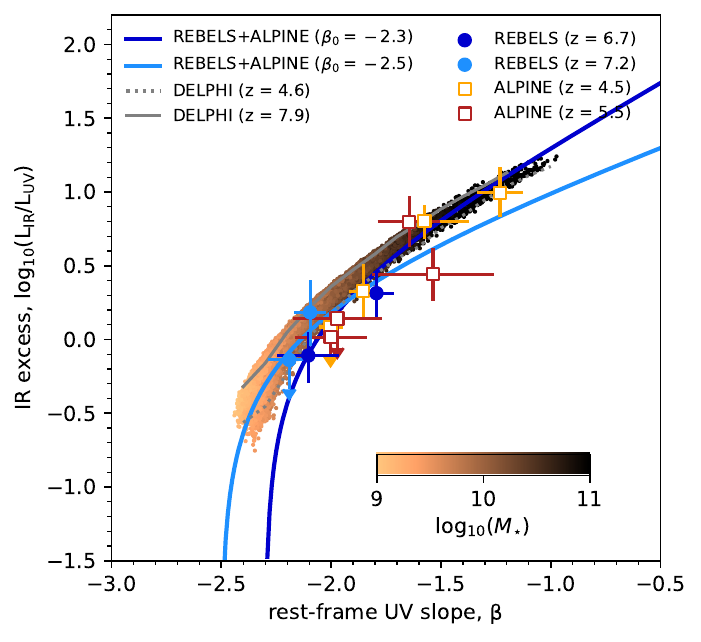}

\caption{The predictions of the DELPHI semi-analytic model of~\citet{Mauerhofer23} for the~\irxb~relation, in comparison to our results from REBELS and ALPINE.
The data points and lines are as shown in Fig.~\ref{fig:irxalpine}.
The observed points from ALPINE that are found at $\beta > -1.75$ correspond to the \lmstar $ = 10$--$11$ stacks of these data.
The REBELS sources show \lmstar $ \simeq 9.5$ (Table~\ref{table:sed}).
The results from DELPHI from $z = 4.6$ to $z = 7.9$ are shown as the small dots, with the colour scale corresponding to \lmstar~as derived in the model.
There is a slight evolution within the redshift range, which we highlight with the solid and dotted grey lines from $z = 7.9$ and $z = 4.6$ respectively.
}\label{fig:delphi}
\end{figure}

The attenuation curve and the~\irxb~relation have been the subject of many theoretical and simulation analyses (e.g. see recent review by~\citealp{Salim20}).
Here we discuss the handful of results that focus specifically on predictions for the high-redshift Universe.
In a work that was based on the REBELS sample,~
\citet{Dayal22} provided the first theoretical comparison to the dust properties of the REBELS galaxies.
Using the semi-analytic model (SAM) {\sc DELPHI}, they were able to match the observed dust masses while also reconciling the intrinsic UV luminosity function with the observed (attenuated) luminosity function at $z = 7$.
\citet{Ferrara22} further identified that some REBELS sources show inconsistencies between the rest-frame UV properties ($\beta$) and FIR emission, suggestive of spatially offset regions within the galaxies.
We recover this result in the form of sources that lie above the typical Calzetti-like \irxb~relation  - showing bluer than expected colours for their measured \Lir. 
While~\citet{Ferrara22} argue that this makes the~\irxb~relation difficult to utilize for these galaxies, in this work we have found that on~\emph{average} there is a correlation in the~\irxb~plane for galaxies in the REBELS sample (that is consistent with a Calzetti-like relation).
In Fig.~\ref{fig:delphi} we present the most recent analysis of dust emission from the DELPHI SAM in comparison to our derived~\irxb~results from REBELS and ALPINE.
We took the output of the model presented in~\citet{Mauerhofer23} and computed the rest-frame UV slope using an identical method to that used on the ALPINE and REBELS stacks, by fitting a power law to the resulting predicted SED.
Sources were extracted that had $-24 <$ \muv $ <-20$, and nebular continuum emission is included in the modelling of the SED.
As is evident from the figure, comparing the DELPHI model of~\citet{Mauerhofer23} to the results of this work we find a good agreement with the model predictions.
The predicted change in IRX and $\beta$ with increasing stellar mass matches very well with what we find in the data, with the observed galaxies in ALPINE at \lmstar $> 10$ lying on top of the model predictions for this mass range.
Furthermore, the bluest $\beta$ values are comparable to those found in the data for galaxies at \lmstar $\simeq 9$, and the lack of evolution we see between $z = 4$--$8$ is also recovered in DELPHI.
As shown in Fig.~\ref{fig:delphi}, the SAM predicts that the~\irxb~should increase with redshift, due to the increased optical depth within compact galaxies at the highest redshifts.
While there are assumptions made in the computation of the~\irxb~in both the models (e.g. through escape fraction of UV photons, dust heating etc.) and in the observations (with the systematic uncertainties in the \Lir determination), it is reassuring that the general trends are present in this basic comparison.

Turning now to works that consider hydrodynamical simulations,~\citet{Narayanan18} used the MAFUSA model to predict the attenuation law at $z = 6$.
They found a relatively grey attenuation curve, with the scatter between different galaxies becoming reduced at higher redshift due to more consistent stellar ages between sources and an increase in complexity in the star-dust geometry.
The predicted curve is very close to that of the~\citet{Calzetti00} curve, however with a prominent 2175\AA~bump.
~\citet{Vijayan23} used the FLARES simulation to investigate the expected dust attenuation curve for galaxies at $z = 5$--$10$, with a particular focus on examining the effect of star-dust geometry.
For galaxies with a similar SFR to the REBELS and ALPINE sources, they find a dust law with a similar slope to that of the~\citet{Calzetti00} relation, despite the input of an SMC-like~\emph{extinction} law to their simulations.
In FLARES, galaxies with a higher SFR show a larger degree of clumpiness, which in turn leads to a larger range in $A_{\rm V}$ over the galaxy surface and hence a greyer attenuation curve when integrated across the source.
The zoom-in cosmological SERRA simulations presented in~\citet{Pallottini22} produces galaxies at $z \simeq 7.7$ that are in excess of the majority of observations at high redshift (and in excess of the Calzetti-relation).
~\citet{Pallottini22} attribute the differences to first, the uncertain FIR SED and hence systemic errors in the derived \Lir~from observations (they suggest that a higher dust temperature should be assumed, exceeding $90\,{\rm K}$), and second, to potential inaccuracies in the feedback prescription in the modelling and an insufficient spatial resolution to fully resolve the molecular clouds where the majority of the FIR luminosity is produced.
Finally,~\citet{Liang19} used the cosmological simulation {\sc MassiveFIRE} to model the~\irxb~relation of $z = 2$--$6$ galaxies.
They found a relation similar to that found locally in the Calzetti-relation (although as a Milky Way dust attenuation curve was input, this is perhaps not surprising)
\citet{Liang19} directly predict the \irxb~values for the galaxies in their simulations, finding objects that are blue ($\beta < -1.5$) and occupy the lower left region of the diagram.
They conclude the scatter around the relation is dominated by different intrinsic $\beta_{0}$ slopes.
In conclusion, in general these models predict a dust attenuation law that is similar to the local~\citet{Calzetti00} relationship for galaxies at $z > 6$ with similar stellar masses and SFRs to the REBELS sample.
When the models have predicted the~\irxb~relation this has been in agreement, or in excess of, what we observe for our combined ALPINE and REBELS samples~\citep{Liang19, Pallottini22, Mauerhofer23}.
We therefore conclude that current models and observational constraints from this work favour a distribution in the~\irxb~plane that is consistent with predictions for a Calzetti-like dust attenuation law at $z = 4$--$8$, with no evidence for a deficit from the local relation towards what would be predicted by a screen of SMC-like dust.

\section{Conclusions}\label{sect:conc}
In this study we present an analysis of the ALMA large program REBELS, which provided observations of the dust continuum emission for a sample of 49 galaxies at $z = 6.5$--$8$.
We also perform a consistent analysis of the ALPINE large program that targeted galaxies in the redshift range of $z = 4$--$6$, to provide a key comparison to the REBELS results and investigate any evolution in the dust emission properties with redshift.
The main conclusions of this work are as follows:
\begin{itemize}
    \item When compared to the expected colour from the extrapolated colour-magnitude relation for fainter galaxies, the REBELS sources are bluer than expected by up to $\Delta \beta = 0.5$--$1.0$ (depending on the extrapolated relation).
    These results point to a flattening, or potential turnover, of the colour magnitude relation bright-ward of \muv $= -21.5$, that can be understand as a consequence of scatter on an underlying steep galaxy stellar mass function.  In this scenario, the REBELS galaxies represent the sub-set of sources at a given \mstar~that have low dust attenuation and hence have bright rest-frame UV magnitudes and a blue colour.
    
    \item When stacking the REBELS sources in the ALMA Band 6 data we find detections at $>5\sigma$ significance for all but the highest redshift and brightest \muv~bins.
    We derive the \Lir~and IRX for the sample from these results, assuming a modified blackbody curve with a dust temperature of $46\,{\rm K}$ and $\beta_{\rm d} = 2.0$ as derived in the model of~\citet{Sommovigo22a}.
    The \irxb~relation we derive for the REBELS sample with this assumed FIR SED is as expected for a Calzetti-like attenuation law (with an intrinsic rest-frame UV slope of $\beta_{0} = -2.3$).
    In comparison to other studies at $z \simeq 7$, we find no strong evidence for a systematic deviation below this relation, although large scatter is found between individual sources.
    
    \item By assuming the same FIR SED for stacks of the ALPINE data (as motivated by the study of~\citealp{Sommovigo22}), we find that the \irxb~results at $z \simeq 4.5$ and $z \simeq 5.5$ produce values that are consistent with our REBELS findings.
    We therefore find negligible evolution in the \irxb~relation from $z = 4$--$8$, and conclude that there is little evidence for a deficit in the \irxb~relation at $ z > 4$ as compared to the local starburst results of e.g.~\citet{Calzetti00}.
    Comparisons to previous studies at $z > 4$ again highlight that the assumed FIR SED has a dramatic affect on the derived \Lir and care must be taken in comparing different studies.
     
    \item We compute the \irxm~relation from our combined ALPINE and REBELS samples, finding a similar slope to previous analyses at $z = 3$, but with a $0.5\,{\rm dex}$ offset to lower IRX at a given \mstar for our assumed FIR SED.  These results corroborate previous findings (e.g.~\citealp{Schouws22, Algera23}) that show that for a given stellar mass, the proportion of obscured star-formation is reduced at $z > 4$ by a factor of $\gtrsim 3$.

    \item In comparison to models of dust attenuation in $z \gtrsim 6$ galaxies, we find that in general a Calzetti-like attenuation curve is predicted.  In simulations that compute the \irxb~relation we find that the results are in good agreement with our combined ALPINE and REBELS analysis, with few studies predicting very low `SMC-like' relations.
    In a detailed comparison to the predictions of the DELPHI SAM, we find very good agreement between this model and our observations over the stellar mass range of \lmstar $= 9$--$11$.
    These results indicate that despite complexities and caveats in both the modelling and observation of high-redshift galaxies, qualitatively (and quantitively for the DELPHI model) there is good agreement between the predicted and measured trends of rest-frame UV colour, dust attenuated star-formation and stellar mass at $z = 4$--$8$.
    
\end{itemize}

To understand the origin of the effects seen in this work and others, a more detailed view on the measured properties of galaxies (such as more precise \mstar, $\beta$ and gas-phase metallicity) are required.
Cycle 1 \emph{JWST} observations of 12 of the REBELS sample are being obtained as part of the General Observer program PID1626 (PI Stefanon).
This program will target the sources with the NIRSpec Integral Field Unit, providing gas-phase properties from the rest-frame optical emission lines for this unique sample.
The data will also provide refined (and resolved) rest-frame UV slope and \mstar~measurements.
For the ALPINE survey, Cycle 2 observations have been approved for a subset of the most massive 18 galaxies for NIRSpec IFU from GO program 3045 (PI Faisst).
The results of these programs, coupled with ongoing additional follow-up with ALMA to determine the dust temperature (via additional bands) and spatial distribution of the gas, dust and stars (via higher spatial resolution ALMA observations), will provide additional insights into the evolution of the \irxb~and~\irxm~relations at $z = 4$--$8$.

\section*{Acknowledgements}
RAAB acknowledges support from an STFC Ernest Rutherford Fellowship [grant number ST/T003596/1].
RJB acknowledges support from NWO grants 600.065.140.11N211 (vrijcompetitie) and TOP grant TOP1.16.057.
RS acknowledges support from an STFC Ernest Rutherford Fellowship [grant number ST/S004831/1].
YF acknowledge support from NAOJ ALMA Scientific Research Grant number 2020-16B. YF further acknowledges support from support from JSPS KAKENHI Grant Number JP23K13149.
MA acknowledges support from FONDECYT grant 1211951, ANID+PCI+REDES 190194 and ANID BASAL project FB210003.
FC acknowledges support from a UKRI Frontier Research Guarantee Grant [grant reference EP/X021025/1]. JSD acknowledges the support of the Royal Society through a Royal Society University Research Professorship.
This work was supported by NAOJ ALMA Scientific Research Grant Code 2021-19A (HI and HSBA). 
HI acknowledges support from JSPS KAKENHI Grant Number JP19K23462.
IDL and MP acknowledge support from ERC starting grant 851622 DustOrigin.
MS acknowledges support from the ERC Consolidator Grant 101088789 (SFEER), from the CIDEGENT/2021/059 grant, and from project PID2019-109592GB-I00/AEI/10.13039/501100011033 from the Spanish Ministerio de Ciencia e Innovaci\'on - Agencia Estatal de Investigaci\'on.
JH acknowledges support of the ERC Consolidator Grant 101088676 (VOYAJ) and the VIDI research programme with project number 639.042.611, which is (partly) financed by the Netherlands Organisation for Scientific Research (NWO).
EdC gratefully acknowledges the Australian Research Council as the recipient of a Future Fellowship (project FT150100079) and the ARC Centre of Excellence for All Sky Astrophysics in 3 Dimensions (ASTRO 3D; project CE170100013). 
This paper makes use of the following ALMA data: ADS/JAO.ALMA\#2019.1.01634.L, ADS/JAO.ALMA\#2017.1.01217.S, ADS/JAO.ALMA\#2017.1.00604.S, ADS/JAO.ALMA\#2018.1.00236.S, ADS/JAO.ALMA\#2018.1.00085.S ADS/JAO.ALMA\#2018.A.00022.S. ALMA is a partnership of ESO (representing its member states), NSF (USA) and NINS (Japan), together with NRC (Canada), MOST and ASIAA (Taiwan), and KASI (Republic of Korea), in cooperation with the Republic of Chile. The Joint ALMA Observatory is operated by ESO, AUI/NRAO and NAOJ. 
PD \& VM acknowledge support from the NWO grant 016.VIDI.189.162 (``ODIN"). PD warmly acknowledges support from the European Commission's and University of Groningen's CO-FUND Rosalind Franklin program.

\section*{Data Availability}
The REBELS and ALPINE datasets used in this work have been publicly released, as have all of the optical and NIR imaging utilized.
The COSMOS2020 catalogue is public~\citep{Weaver22}.




\bibliographystyle{mnras}
\bibliography{bibtex_parsedrebels} 




\appendix

\section{ALPINE analysis}\label{sect:appendalpine}

\begin{figure}

\includegraphics[width = 0.45\textwidth]{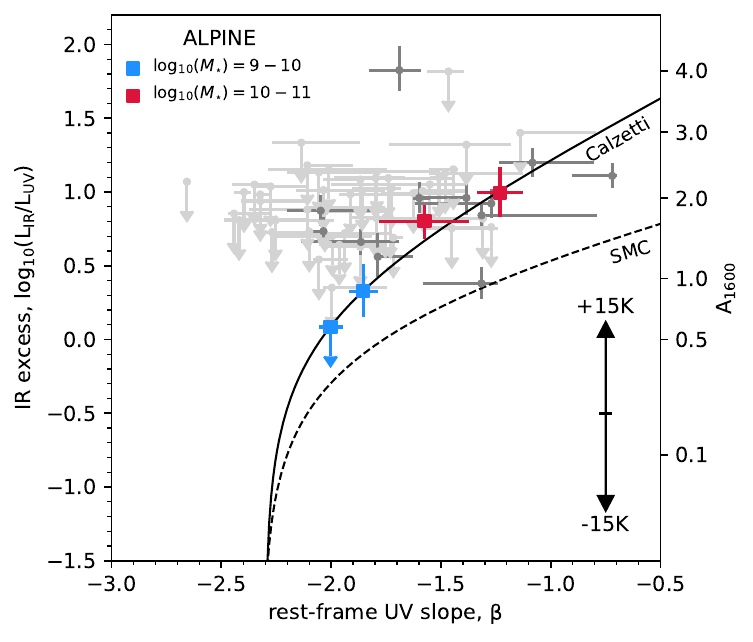}\\
\includegraphics[width = 0.45\textwidth]{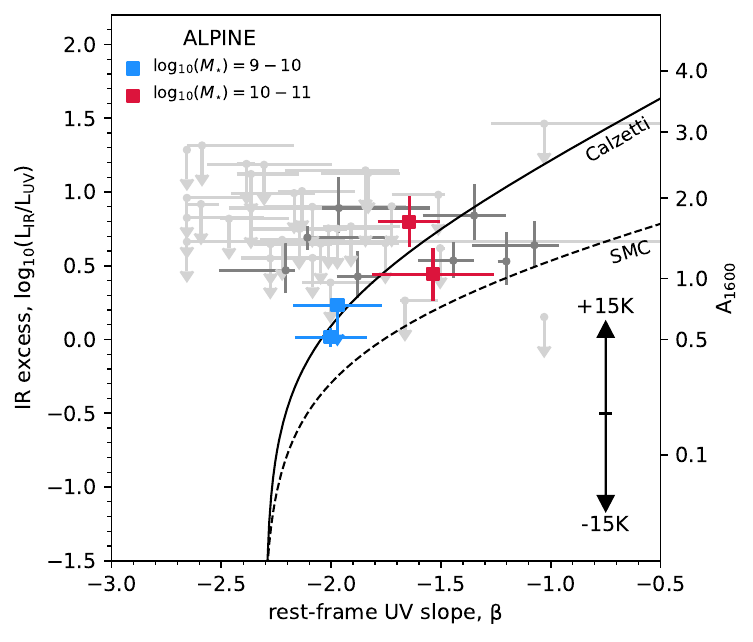}

\caption{The results of our reanalysis of the ALPINE dataset on the~\irxb~relation, with the upper (lower) plot showing the $z \simeq 4.5$ ($z \simeq 5.5$) results.
The individual galaxy values where there is a detection in the dust continuum are shown as the filled grey points, with undetected sources shown as the light grey limits.
The results of our stacking analysis, which takes into account the extended dust emission observed for the stacks at ${\rm log}_{10}(M_{\star}/{\rm M}_{\odot}) > 10$, are shown as the blue and red points.  
The Calzetti and SMC-like relations are shown as described as in Fig.~\ref{fig:irx}.
}\label{fig:irxalpine}
\end{figure}

We show the results of our analysis of the ALMA sample for the two redshift ranges centred on $z \simeq 4.5$ and $z \simeq 5.5$ in Fig.~\ref{fig:irxalpine}.
The data are tabulated in Table~\ref{tab:alpine_flux}.
The individual points are plotted using the $\beta$-values as derived by fitting to the best-fit SED by~\citet{Faisst20} with some objects hitting the edge of parameter space, while the IRX value was recomputed in this work.
The errors on the individual $\beta$-slopes are smaller in ALPINE than in REBELS, due to the sources being in general brighter and having more photometric bands covering the rest-frame UV part of the SED.
There are a larger fraction of non-detections in the dust continuum in ALPINE (80 percent;~\citealp{Fudamoto20}) in comparison to REBELS (63 percent;~\citealp{Inami22}).
From the bootstrap resampling used to derive errors on the stacked ALMA flux, we also find evidence for an increased scatter within the bins than found in the REBELS sample.

\begin{table*}
\caption{The measured FIR fluxes and derived properties of the eight ALPINE stacks.
The top (bottom) two rows show the lower (higher) redshift stack, with the rows ordered by $M_{\rm UV}$ bin.
Columns 1 and 2 detail the redshift and number of sources included in each stack.
The average redshift and $M_{\rm UV}$ of each stack are shown in Columns 3 and 4.
Column 5 presents the measured peak ALMA flux, with the corresponding S/N shown in brackets.
The flux measured using a Gaussian fit is shown in Column 6.
The derived FIR luminosity (assuming a $T_{\rm dust} = 46\,{\rm K}$ and $\beta_{\rm d} = 2.0$) and the resulting IRX values are shown in Columns 7 and 8.
The \Lir~was determined from the peak flux for \lmstar $< 10$ and the Gaussian flux at higher masses, as the \lmstar $= 10$--$11$ stack shows evidence for being spatially extended.
Finally the rest-frame UV slope $\beta$ is presented in Column 9, as measured from the best-fitting SED model.
}
    \centering
   \begin{tabular}{cccccccccc}
\hline
Mass bin & $M_{\rm UV}$ bin & N & $z_{\rm mean}$ & $M_{\rm UV}$ & $F_{\rm Peak}$ & $F_{\rm Gauss}$ & \Lir & IRX &
$\beta$ \\
& & & & $/{\rm mag}$ & $/\mu\,{\rm Jy}$ & $/\mu\,{\rm Jy}$  & $/10^{11}\,{\rm L}_{\odot}$ & \\
\hline
9--10 & $-23.5< M_{\rm UV} < -22.0$ & 7 & 4.53 & $-22.20 \pm 0.20$ & $ < 84.0\,(3.3)$ & $--$ & $ < 2.0 $ & $ < 0.08 $ & $ -2.00_{-0.05}^{+0.05}$ \\[1ex]
9--10 & $-22.0< M_{\rm UV} < -20.5$ & 34 & 4.52 & $-21.31 \pm 0.25$ & $ 64.4 \pm 11.6 \, (7.0)$ & $ 111.3 \pm 23.4 $ & $ 1.5 \pm 0.3$ & $ 0.33_{-0.18}^{+0.19}$ & $ -1.85_{-0.07}^{+0.07}$ \\[1ex]
10--11 & $-23.5< M_{\rm UV} < -22.0$ & 7 & 4.52 & $-22.15 \pm 0.09$ & $ 159.4 \pm 55.9 \, (7.6)$ & $ 422.8 \pm 74.6 $ & $ 9.7 \pm 1.7$ & $ 0.80_{-0.12}^{+0.11}$ & $ -1.58_{-0.20}^{+0.20}$ \\[1ex]
10--11 & $-22.0< M_{\rm UV} < -20.5$ & 6 & 4.52 & $-21.31 \pm 0.25$ & $ 202.1 \pm 108.7 \, (10.1)$ & $ 299.6 \pm 45.6 $ & $ 7.0 \pm 1.1$ & $ 0.99_{-0.16
}^{+0.18}$ & $ -1.23_{-0.10}^{+0.10}$ \\

\hline

9--10 & $-23.5< M_{\rm UV} < -22.0$ & 5 & 5.43 & $-22.38 \pm 0.07$ & $ < 85.2\,(2.3)$ & $--$ & $ < 2.6 $ & $ < 0.14 $ & $ -1.97_{-0.20}^{+0.20}$ \\[1ex]
9--10 & $-22.0< M_{\rm UV} < -20.5$ & 18 & 5.57 & $-21.50 \pm 0.20$ & $ 27.1 \pm 5.6 \, (3.4)$ & $ 40.1 \pm 18.1 $ & $ 0.9 \pm 0.2$ & $ 0.02_{-0.18}^{+0.17}$ & $ -2.00_{-0.16}^{+0.16}$ \\[1ex]
10--11 & $-23.5< M_{\rm UV} < -22.0$ & 5 & 5.47 & $-22.53 \pm 0.22$ & $ 98.0 \pm 26.3 \, (7.0)$ & $ 192.1 \pm 39.1 $ & $ 6.0 \pm 1.2$ & $ 0.44_{-0.18}^{
+0.18}$ & $ -1.54_{-0.28}^{+0.28}$ \\[1ex]
10--11 & $-22.0< M_{\rm UV} < -20.5$ & 4 & 5.57 & $-21.74 \pm 0.23$ & $ 122.0 \pm 29.6 \, (7.9)$ & $ 202.9 \pm 38.1 $ & $ 6.6 \pm 1.2$ & $ 0.80_{-0.17}^
{+0.18}$ & $ -1.65_{-0.14}^{+0.14}$ \\

\hline
\end{tabular}\label{tab:alpine_flux}
\end{table*}

\section{REBELS individual values}

In Table~\ref{table:indi} (continued in~\ref{table:indicont}) we present the properties of the individual galaxies in the REBELS sample, as derived for our~\irxb~and~\irxm~analyses.

\begin{table*}\caption{The observed and derived properties for the individual galaxies in the REBELS sample.
Column 1, 2 and 3 give the REBELS-ID, redshift and stellar mass as presented in~\citet{Bouwens22}.
A redshift value marked with an asterisk indicates a spectroscopic redshift derived from~\cii.
The rest-frame UV slope, $\beta$, and the \muv~as derived in Stefanon et al. (in prep) are shown in Columns 4 and 5.
Column 6 and 7 present the \Lir~and IRX computed in this work assuming a $T_{\rm d} = 46\,{\rm K}$ and $\beta_{\rm d} = 2.0$.
The final three columns present the SFR derived from the rest-frame UV and FIR fluxes via the calibrations presented in Section~\ref{sect:sfr}, and the fraction of obscured star-formation this corresponds to.
}\label{table:indi}
\begin{tabular}{cccccccccc}
\hline
ID & $z$ & \lmstar & $\beta$ & \muv & \Lir & ${\rm IRX}$ & ${\rm SFR}_{\rm UV}$ & ${\rm SFR}_{\rm IR}$ & $f_{\rm obs}$ \\
 & &  & & $/{\rm mag}$ & $/10^{11}\,{\rm L}_{\odot}$ & \sfrunit & \sfrunit & \\
\hline
REBELS-01 & 7.18* & $10.0_{-0.5}^{+0.4}$ & $ -2.04_{-0.20}^{+0.24}$ & $ -22.87_{-0.11}^{+0.11}$ & $ < 2.8 $ & $ < -0.03 $ & $ 38_{-4}^{+4}$ & $ < 33 $ & $ < 0.47 $ \\[1ex]
REBELS-02 & 6.64 & $9.0_{-0.5}^{+0.5}$ & $ -2.24_{-0.36}^{+0.44}$ & $ -22.11_{-0.19}^{+0.19}$ & $ < 2.3 $ & $ < 0.18 $ & $ 19_{-3}^{+4}$ & $ < 27 $ & $ < 0.59 $ \\[1ex]
REBELS-03 & 6.97* & $9.1_{-0.9}^{+0.6}$ & $ -2.14_{-0.46}^{+0.63}$ & $ -21.81_{-0.26}^{+0.26}$ & $ < 2.6 $ & $ < 0.36 $ & $ 14_{-3}^{+4}$ & $ < 31 $ & $ < 0.68 $ \\[1ex]
REBELS-04 & 8.57 & $8.7_{-0.7}^{+1.0}$ & $ -2.15_{-0.38}^{+0.20}$ & $ -22.33_{-0.03}^{+0.03}$ & $ 1.3 \pm 0.4$ & $ -0.15_{-0.17}^{+0.13}$ & $ 23_{-1}^{+1}$ & $ 15_{-5}^{+
5}$ & $ 0.40 \pm 0.12$ \\[1ex]
REBELS-05 & 6.50* & $9.2_{-1.0}^{+0.8}$ & $ -1.29_{-0.44}^{+0.36}$ & $ -21.57_{-0.23}^{+0.23}$ & $ 3.2 \pm 0.6$ & $ 0.55_{-0.17}^{+0.18}$ & $ 11_{-2}^{+3}$ & $ 38_{-7}^{+
7}$ & $ 0.77 \pm 0.22$ \\[1ex]
REBELS-06 & 6.80 & $9.5_{-0.8}^{+0.5}$ & $ -1.24_{-0.35}^{+0.67}$ & $ -21.73_{-0.28}^{+0.28}$ & $ 3.7 \pm 0.7$ & $ 0.54_{-0.20}^{+0.21}$ & $ 13_{-3}^{+4}$ & $ 44_{-9}^{+9
}$ & $ 0.77 \pm 0.25$ \\[1ex]
REBELS-07 & 7.15 & $8.7_{-0.8}^{+0.7}$ & $ -2.39_{-0.43}^{+0.37}$ & $ -22.06_{-0.27}^{+0.27}$ & $ < 2.6 $ & $ < 0.27 $ & $ 18_{-4}^{+5}$ & $ < 31 $ & $ < 0.64 $ \\[1ex]
REBELS-08 & 6.75* & $9.0_{-0.7}^{+0.6}$ & $ -2.17_{-0.58}^{+0.58}$ & $ -21.82_{-0.40}^{+0.40}$ & $ 4.7 \pm 0.9$ & $ 0.62_{-0.23}^{+0.28}$ & $ 14_{-4}^{+6}$ & $ 57_{-11}^{
+11}$ & $ 0.80 \pm 0.34$ \\[1ex]
REBELS-09 & 7.61 & $8.7_{-0.3}^{+0.3}$ & $ -2.66_{-0.53}^{+0.93}$ & $ -23.03_{-0.31}^{+0.31}$ & $ < 3.0 $ & $ < -0.06 $ & $ 44_{-11}^{+15}$ & $ < 36 $ & $ < 0.45 $ \\[1ex]
REBELS-10 & 7.42 & $10.2_{-0.4}^{+0.3}$ & $ -1.34_{-0.83}^{+0.48}$ & $ -22.69_{-0.30}^{+0.30}$ & $ < 3.2 $ & $ < 0.10 $ & $ 32_{-8}^{+10}$ & $ < 38 $ & $ < 0.54 $ \\[1ex]
REBELS-11 & 8.24 & $9.4_{-0.8}^{+0.7}$ & $ -1.60_{-1.15}^{+0.17}$ & $ -22.77_{-0.22}^{+0.22}$ & $ < 1.7 $ & $ < -0.19 $ & $ 35_{-6}^{+8}$ & $ < 21 $ & $ < 0.38 $ \\[1ex]
REBELS-12 & 7.35* & $8.9_{-0.7}^{+0.9}$ & $ -1.99_{-0.76}^{+0.48}$ & $ -22.46_{-0.30}^{+0.30}$ & $ 4.7 \pm 1.3$ & $ 0.37_{-0.25}^{+0.25}$ & $ 26_{-6}^{+8}$ & $ 57_{-16}^{
+16}$ & $ 0.69 \pm 0.27$ \\[1ex]
REBELS-13 & 8.19 & $9.8_{-0.5}^{+0.5}$ & $ -1.08_{-0.65}^{+0.59}$ & $ -22.90_{-0.23}^{+0.23}$ & $ < 1.6 $ & $ < -0.27 $ & $ 39_{-7}^{+9}$ & $ < 20 $ & $ < 0.33 $ \\[1ex]
REBELS-14 & 7.08* & $8.7_{-0.7}^{+0.8}$ & $ -2.21_{-0.47}^{+0.41}$ & $ -22.66_{-0.39}^{+0.39}$ & $ 3.0 \pm 0.7$ & $ 0.08_{-0.26}^{+0.29}$ & $ 31_{-9}^{+14}$ & $ 35_{-9}^{
+9}$ & $ 0.53 \pm 0.23$ \\[1ex]
REBELS-15 & 6.88* & $8.8_{-0.5}^{+0.5}$ & $ -2.18_{-0.50}^{+0.52}$ & $ -22.58_{-0.31}^{+0.31}$ & $ < 3.2 $ & $ < 0.15 $ & $ 29_{-7}^{+10}$ & $ < 38 $ & $ < 0.57 $ \\[1ex]
REBELS-16 & 6.70 & $9.9_{-0.4}^{+0.3}$ & $ -1.70_{-0.76}^{+0.48}$ & $ -22.47_{-0.05}^{+0.05}$ & $ < 3.4 $ & $ < 0.21 $ & $ 26_{-1}^{+1}$ & $ < 40 $ & $ < 0.61 $ \\[1ex]
REBELS-17 & 6.54* & $9.1_{-0.6}^{+0.6}$ & $ -1.70_{-0.47}^{+0.33}$ & $ -21.64_{-0.24}^{+0.24}$ & $ < 3.8 $ & $ < 0.59 $ & $ 12_{-2}^{+3}$ & $ < 45 $ & $ < 0.79 $ \\[1ex]
REBELS-18 & 7.68* & $9.5_{-0.7}^{+0.6}$ & $ -1.34_{-0.32}^{+0.19}$ & $ -22.37_{-0.15}^{+0.15}$ & $ 3.9 \pm 0.7$ & $ 0.31_{-0.15}^{+0.14}$ & $ 24_{-3}^{+4}$ & $ 46_{-9}^{+
9}$ & $ 0.66 \pm 0.15$ \\[1ex]
REBELS-19 & 7.37* & $8.8_{-0.7}^{+0.7}$ & $ -2.33_{-0.64}^{+0.45}$ & $ -21.60_{-0.24}^{+0.24}$ & $ 3.9 \pm 1.1$ & $ 0.63_{-0.23}^{+0.22}$ & $ 12_{-2}^{+3}$ & $ 47_{-13}^{
+13}$ & $ 0.80 \pm 0.29$ \\[1ex]
REBELS-20 & 7.12 & $8.6_{-0.6}^{+0.6}$ & $ -2.59_{-0.60}^{+0.57}$ & $ -21.85_{-0.11}^{+0.11}$ & $ < 4.0 $ & $ < 0.53 $ & $ 15_{-1}^{+2}$ & $ < 47 $ & $ < 0.76 $ \\[1ex]
REBELS-21 & 6.59 & $10.4_{-0.4}^{+0.2}$ & $ -2.15_{-0.24}^{+0.42}$ & $ -21.91_{-0.21}^{+0.21}$ & $ < 2.7 $ & $ < 0.34 $ & $ 16_{-3}^{+3}$ & $ < 32 $ & $ < 0.67 $ \\[1ex]
REBELS-22 & 7.48 & $9.7_{-0.8}^{+0.4}$ & $ -2.23_{-0.30}^{+0.21}$ & $ -22.26_{-0.11}^{+0.11}$ & $ < 2.5 $ & $ < 0.16 $ & $ 22_{-2}^{+2}$ & $ < 30 $ & $ < 0.58 $ \\[1ex]
REBELS-23 & 6.64* & $9.1_{-0.6}^{+0.5}$ & $ -1.57_{-0.45}^{+0.28}$ & $ -21.64_{-0.52}^{+0.52}$ & $ < 3.6 $ & $ < 0.58 $ & $ 12_{-5}^{+8}$ & $ < 44 $ & $ < 0.78 $ \\[1ex]
REBELS-24 & 8.35 & $9.0_{-0.9}^{+0.9}$ & $ -1.56_{-0.83}^{+0.56}$ & $ -22.04_{-0.24}^{+0.24}$ & $ < 3.0 $ & $ < 0.34 $ & $ 18_{-4}^{+4}$ & $ < 36 $ & $ < 0.67 $ \\[1ex]
REBELS-25 & 7.31* & $9.9_{-0.2}^{+0.1}$ & $ -1.85_{-0.46}^{+0.56}$ & $ -21.67_{-0.23}^{+0.23}$ & $ 14.4 \pm 1.2$ & $ 1.16_{-0.12}^{+0.14}$ & $ 13_{-2}^{+3}$ & $ 173_{-15}
^{+15}$ & $ 0.93 \pm 0.21$ \\[1ex]
REBELS-26 & 6.60* & $9.5_{-0.8}^{+0.5}$ & $ -1.92_{-0.25}^{+0.19}$ & $ -21.84_{-0.12}^{+0.12}$ & $ < 4.5 $ & $ < 0.59 $ & $ 15_{-2}^{+2}$ & $ < 54 $ & $ < 0.79 $ \\[1ex]
REBELS-27 & 7.09* & $9.7_{-0.3}^{+0.2}$ & $ -1.79_{-0.45}^{+0.42}$ & $ -21.93_{-0.24}^{+0.24}$ & $ 2.9 \pm 0.6$ & $ 0.36_{-0.18}^{+0.19}$ & $ 16_{-3}^{+4}$ & $ 34_{-7}^{+
7}$ & $ 0.68 \pm 0.20$ \\[1ex]
REBELS-28 & 6.94* & $8.6_{-0.5}^{+0.7}$ & $ -1.95_{-0.36}^{+0.29}$ & $ -22.46_{-0.29}^{+0.29}$ & $ < 3.0 $ & $ < 0.17 $ & $ 26_{-6}^{+8}$ & $ < 36 $ & $ < 0.58 $ \\[1ex]
REBELS-29 & 6.68* & $9.6_{-0.2}^{+0.2}$ & $ -1.61_{-0.19}^{+0.10}$ & $ -22.24_{-0.12}^{+0.12}$ & $ 2.8 \pm 0.6$ & $ 0.23_{-0.16}^{+0.14}$ & $ 21_{-2}^{+2}$ & $ 34_{-8}^{+
8}$ & $ 0.61 \pm 0.16$ \\[1ex]
REBELS-30 & 6.98* & $9.3_{-0.6}^{+0.5}$ & $ -1.95_{-0.22}^{+0.15}$ & $ -22.35_{-0.09}^{+0.09}$ & $ < 2.6 $ & $ < 0.15 $ & $ 24_{-2}^{+2}$ & $ < 31 $ & $ < 0.57 $ \\[1ex]
REBELS-31 & 6.68 & $9.2_{-0.3}^{+0.3}$ & $ -2.27_{-0.33}^{+0.18}$ & $ -22.30_{-0.18}^{+0.18}$ & $ < 3.9 $ & $ < 0.34 $ & $ 22_{-3}^{+4}$ & $ < 46 $ & $ < 0.67 $ \\[1ex]
REBELS-32 & 6.73* & $9.6_{-0.4}^{+0.4}$ & $ -1.50_{-0.30}^{+0.28}$ & $ -21.65_{-0.14}^{+0.14}$ & $ 2.9 \pm 0.8$ & $ 0.48_{-0.20}^{+0.17}$ & $ 12_{-1}^{+2}$ & $ 35_{-10}^{
+10}$ & $ 0.74 \pm 0.23$ \\[1ex]
REBELS-33 & 6.67 & $9.4_{-0.5}^{+0.4}$ & $ -2.04_{-0.71}^{+0.24}$ & $ -21.58_{-0.15}^{+0.15}$ & $ < 3.7 $ & $ < 0.61 $ & $ 12_{-1}^{+2}$ & $ < 44 $ & $ < 0.79 $ \\[1ex]
REBELS-34 & 6.63* & $9.3_{-0.3}^{+0.3}$ & $ -2.02_{-0.15}^{+0.07}$ & $ -22.47_{-0.08}^{+0.08}$ & $ < 3.5 $ & $ < 0.23 $ & $ 26_{-2}^{+2}$ & $ < 42 $ & $ < 0.62 $ \\[1ex]
REBELS-35 & 6.97 & $8.9_{-0.7}^{+0.7}$ & $ -2.07_{-1.12}^{+0.27}$ & $ -22.51_{-0.11}^{+0.11}$ & $ < 3.5 $ & $ < 0.21 $ & $ 27_{-3}^{+3}$ & $ < 42 $ & $ < 0.60 $ \\[1ex]
REBELS-36 & 7.68* & $9.4_{-0.9}^{+0.8}$ & $ -2.57_{-0.47}^{+0.48}$ & $ -22.20_{-0.20}^{+0.20}$ & $ < 2.9 $ & $ < 0.25 $ & $ 21_{-3}^{+4}$ & $ < 34 $ & $ < 0.63 $ \\[1ex]
REBELS-37 & 7.75 & $8.6_{-0.7}^{+0.7}$ & $ -1.24_{-0.27}^{+0.16}$ & $ -22.25_{-0.03}^{+0.03}$ & $ 1.8 \pm 0.3$ & $ 0.03_{-0.09}^{+0.08}$ & $ 21_{-1}^{+1}$ & $ 22_{-4}^{+4
}$ & $ 0.50 \pm 0.08$ \\[1ex]
REBELS-38 & 6.58* & $9.6_{-1.3}^{+0.7}$ & $ -2.18_{-0.42}^{+0.45}$ & $ -21.87_{-0.25}^{+0.25}$ & $ 7.7 \pm 1.1$ & $ 0.81_{-0.16}^{+0.17}$ & $ 15_{-3}^{+4}$ & $ 93_{-13}^{
+13}$ & $ 0.86 \pm 0.23$ \\[1ex]
REBELS-39 & 6.85* & $8.6_{-0.6}^{+0.6}$ & $ -1.96_{-0.28}^{+0.30}$ & $ -22.71_{-0.17}^{+0.17}$ & $ 3.8 \pm 0.8$ & $ 0.17_{-0.16}^{+0.15}$ & $ 33_{-5}^{+6}$ & $ 45_{-9}^{+
9}$ & $ 0.58 \pm 0.15$ \\[1ex]
REBELS-40 & 7.37* & $9.5_{-1.0}^{+0.5}$ & $ -1.44_{-0.36}^{+0.29}$ & $ -21.84_{-0.07}^{+0.07}$ & $ 2.8 \pm 0.8$ & $ 0.39_{-0.16}^{+0.13}$ & $ 15_{-1}^{+1}$ & $ 34_{-9}^{+
9}$ & $ 0.70 \pm 0.19$ \\
\hline
\end{tabular}
\end{table*}

\begin{table*}\caption{Continued.
}\label{table:indicont}
\begin{tabular}{cccccccccc}
\hline
ID & $z$ & \lmstar & $\beta$ & \muv & \Lir & ${\rm IRX}$ & ${\rm SFR}_{\rm UV}$ & ${\rm SFR}_{\rm IR}$ & $f_{\rm obs}$ \\
 & &  & & $/{\rm mag}$ & $/10^{11}\,{\rm L}_{\odot}$ & \sfrunit & \sfrunit & \\
\hline
REBELS-P1 & 6.75* & $9.5_{-0.4}^{+0.3}$ & $ -2.53_{-0.22}^{+0.14}$ & $ -22.26_{-0.11}^{+0.11}$ & $ < 3.1 $ & $ < 0.27 $ & $ 22_{-2}^{+2}$ & $ < 38 $ & $ < 0.63 $ \\[1ex]
REBELS-P2 & 8.47 & $8.8_{-0.6}^{+0.6}$ & $ -1.85_{-0.24}^{+0.19}$ & $ -21.74_{-0.36}^{+0.36}$ & $ < 3.1 $ & $ < 0.47 $ & $ 13_{-4}^{+5}$ & $ < 38 $ & $ < 0.74 $ \\[1ex]
REBELS-P3 & 7.69 & $8.7_{-0.6}^{+0.5}$ & $ -1.83_{-0.12}^{+0.10}$ & $ -21.66_{-0.34}^{+0.34}$ & $ < 3.4 $ & $ < 0.54 $ & $ 12_{-3}^{+5}$ & $ < 41 $ & $ < 0.77 $ \\[1ex]
REBELS-P4 & 8.32 & $9.2_{-1.0}^{+0.7}$ & $ -2.28_{-0.25}^{+0.26}$ & $ -21.92_{-0.43}^{+0.43}$ & $ < 3.7 $ & $ < 0.48 $ & $ 16_{-5}^{+8}$ & $ < 45 $ & $ < 0.74 $ \\[1ex]
REBELS-P5 & 7.19 & $8.9_{-0.6}^{+0.6}$ & $ -1.69_{-0.33}^{+0.25}$ & $ -22.26_{-0.13}^{+0.13}$ & $ < 3.0 $ & $ < 0.25 $ & $ 22_{-2}^{+3}$ & $ < 37 $ & $ < 0.63 $ \\[1ex]
REBELS-P6 & 6.81* & $9.6_{-0.5}^{+0.5}$ & $ -2.18_{-0.21}^{+0.16}$ & $ -21.75_{-0.12}^{+0.12}$ & $ < 5.3 $ & $ < 0.70 $ & $ 14_{-1}^{+2}$ & $ < 63 $ & $ < 0.82 $ \\[1ex]
REBELS-P7 & 6.75* & $9.8_{-0.3}^{+0.3}$ & $ -2.09_{-0.15}^{+0.14}$ & $ -21.63_{-0.14}^{+0.14}$ & $ 2.9 \pm 0.8$ & $ 0.48_{-0.18}^{+0.16}$ & $ 12_{-1}^{+2}$ & $ 35_{-9}^{+
9}$ & $ 0.74 \pm 0.22$ \\[1ex]
REBELS-P8 & 6.85* & $9.5_{-0.5}^{+0.5}$ & $ -2.49_{-0.62}^{+0.61}$ & $ -21.84_{-0.09}^{+0.09}$ & $ < 5.9 $ & $ < 0.71 $ & $ 15_{-1}^{+1}$ & $ < 70 $ & $ < 0.83 $ \\[1ex]
REBELS-P9 & 7.06* & $9.3_{-0.4}^{+0.4}$ & $ -1.76_{-0.33}^{+0.17}$ & $ -23.00_{-0.07}^{+0.07}$ & $ 2.9 \pm 0.9$ & $ -0.07_{-0.18}^{+0.14}$ & $ 43_{-3}^{+3}$ & $ 34_{-10}^
{+10}$ & $ 0.45 \pm 0.14$ \\[1ex]
\hline
\end{tabular}
\end{table*}

\section*{Affiliations}
\noindent
{\it
$^{1}$Jodrell Bank Centre for Astrophysics, Department of Physics and Astronomy, School of Natural Sciences, The University of Manchester, Manchester, M13 9PL, UK\\
$^{2}$Hiroshima Astrophysical Science Center, Hiroshima University, 1-3-1 Kagamiyama, Higashi-Hiroshima, Hiroshima 739-8526, Japan \\
$^{3}$Scuola Normale Superiore, Piazza dei Cavalieri 7, I-56126 Pisa, Italy \\
$^{4}$Astrophysics Research Institute, Liverpool John Moores University, 146 Brownlow Hill, Liverpool L3 5RF, UK\\
$^{5}$National Astronomical Observatory of Japan, 2-21-1, Osawa, Mitaka, Tokyo 181-8588, Japan\\
$^{6}$ Instituto de Estudios Astrof\'{\i}sicos, Facultad de Ingenier\'{\i}a y Ciencias, Universidad Diego Portales, Av. Ej\'ercito 441, Santiago, Chile\\
$^{7}$Observatoire de Geneve, CH-1290 Versoix, Switzerland\\
$^{8}$Leiden Observatory, Leiden University, NL-2300 RA Leiden, Netherlands \\
$^{9}$International Centre for Radio Astronomy Research, University of Western Australia, 35 Stirling Hwy, Crawley 26WA 6009, Australia\\
$^{10}$ARC Centre of Excellence for All Sky Astrophysics in 3 Dimensions (ASTRO 3D), Australia 
$^{11}$Institute for Astronomy, University of Edinburgh, Royal Observatory, Edinburgh, EH9 3HJ, UK\\
$^{12}$Kapteyn Astronomical Institute , University of Groningen, NL-9700 AV Groningen, the Netherlands\\
$^{13}$Sterrenkundig Observatorium, Ghent University, Krijgslaan 281 - S9, B-9000 Gent, Belgium\\
$^{14}$Dept. of Physics \& Astronomy, University College London, Gower Street, London WC1E 6BT, UK\\
$^{15}$Waseda Research Institute for Science and Engineering, Faculty of Science and Engineering, Waseda University, 3-4-1 Okubo, Shinjuku, Tokyo 169-8555, Japan\\
$^{16}$Departament d’Astronomia i Astrofısica, Universitat de Valencia, C. Dr. Moliner 50, E-46100 Burjassot, Valencia, Spain\\
$^{17}$Dipartimento di Fisica, Sapienza, Universita di Roma, Piazzale Aldo Moro 5, I-00185 Roma, Italy\\
$^{18}$Sapienza School for Advanced Studies, Viale Regina Elena 291, I-00161 Roma Italy\\
$^{19}$INAF/Osservatorio Astronomico di Roma, via Frascati 33, I-00078 Monte Porzio Catone, Roma, Italy\\
$^{20}$Istituto Nazionale di Fisica Nucleare, Sezione di Roma1, Piazzale Aldo Moro 2, I-00185 Roma Italy\\
$^{21}$Centre for Astrophysics \& Supercomputing, Swinburne University of Technology, PO Box 218, Hawthorn, VIC 3112, Australia\\
$^{22}$Steward Observatory, University of Arizona, 933 N Cherry Ave, Tucson, AZ 85721, USA\\

}

\bsp	
\label{lastpage}
\end{document}